\documentclass[twocolumn,twocolappendix]{aastex7}

\usepackage{amsmath}

\usepackage{booktabs}   
\usepackage{multirow} 
\shorttitle{GRB 260310A/SN 2026fgk}
\shortauthors{O'Connor et al.}
\submitjournal{ApJ}

\usepackage{siunitx}
\usepackage{longtable}
\usepackage{caption}

\definecolor{blazeorange}{rgb}{1.0, 0.4, 0.0}
\definecolor{seagreen}{rgb}{0.18, 0.55, 0.34}
\definecolor{darkgreen}{rgb}{0.08, 0.45, 0.2}
\definecolor{rufous}{rgb}{0.66, 0.11, 0.03}
\definecolor{royalfuchsia}{rgb}{0.79, 0.17, 0.57}
\definecolor{scarlet}{rgb}{1.0, 0.13, 0.0}
\definecolor{royalpurple}{rgb}{0.47, 0.32, 0.66}

\begin{document}

\title{
GRB 260310A/SN 2026fgk: Photometric and Spectroscopic Evolution of a Nearby GRB-Supernova and an Exceptionally Bright Afterglow at $z=0.153$
}

\correspondingauthor{Brendan O'Connor}
\author[0000-0002-9700-0036]{Brendan O'Connor}
    \altaffiliation{McWilliams Fellow}
    \affiliation{McWilliams Center for Cosmology and Astrophysics, Department of Physics, Carnegie Mellon University, Pittsburgh, PA 15213, USA}
    \email[show]{boconno2@andrew.cmu.edu}  

\author[0009-0001-0574-2332]{Malte Busmann}
    \affiliation{University Observatory, Faculty of Physics, Ludwig-Maximilians-Universität München, Scheinerstr. 1, 81679 Munich, Germany}
    \affiliation{Excellence Cluster ORIGINS, Boltzmannstr. 2, 85748 Garching, Germany}
    \email{m.busmann@physik.lmu.de}
    
\author[0000-0002-9364-5419]{Xander J. Hall}
\affiliation{McWilliams Center for Cosmology and Astrophysics, Department of Physics, Carnegie Mellon University, Pittsburgh, PA 15213, USA}
\email{xhall@cmu.edu}

\author[0000-0002-8482-8993]{Kenta Taguchi}
    \affiliation{Department of Astronomy, Kyoto University, Kitashirakawa-Oiwake-cho, Sakyo-ku, Kyoto 606-8502, Japan}
    \email{kentagch@kusastro.kyoto-u.ac.jp}

\author[0000-0001-8253-6850]{Masaomi Tanaka}
    \affiliation{Astronomical Institute, Tohoku University, Sendai 980-8578, Japan}
    \affiliation{Division for the Establishment of Frontier Sciences, Organization for Advanced Studies, Tohoku University, Sendai 980-8577, Japan}
    \email{masaomi.tanaka@astr.tohoku.ac.jp}

\author[0000-0003-3270-7644]{Daniel Gruen}
	\affiliation{University Observatory, Faculty of Physics, Ludwig-Maximilians-Universität München, Scheinerstr. 1, 81679 Munich, Germany}
	\affiliation{Excellence Cluster ORIGINS, Boltzmannstr. 2, 85748 Garching, Germany}
	\email{daniel.gruen@lmu.de}

\author[0009-0002-5156-7819]{Seiji Toshikage}
    \affiliation{Astronomical Institute, Tohoku University, Sendai 980-8578, Japan}
    \email{seiji.toshikage@astr.tohoku.ac.jp}

\author[0000-0003-3433-2698]{Ariel J. Amsellem}
    \affiliation{McWilliams Center for Cosmology and Astrophysics, Department of Physics, Carnegie Mellon University, Pittsburgh, PA 15213, USA}
    \email{aamselle@andrew.cmu.edu}

\author[0009-0001-8023-5701]{Ziyuan Zhu}
	\affiliation{University Observatory, Faculty of Physics, Ludwig-Maximilians-Universität München, Scheinerstr. 1, 81679 Munich, Germany}
    \email{ziyuan.zhu@campus.lmu.de}

\author[0000-0002-6011-0530]{Antonella Palmese}
\affiliation{McWilliams Center for Cosmology and Astrophysics, Department of Physics, Carnegie Mellon University, Pittsburgh, PA 15213, USA}
\email{palmese@cmu.edu}

\author[0000-0002-0676-3661]{Dylan Green}
    \affiliation{Lawrence Berkeley National Laboratory, 1 Cyclotron Road, Berkeley, CA 94720, USA}
    \email{dylangreen@lbl.gov}

\author[0000-0003-0776-8859]{John Banovetz}
    \affiliation{Lawrence Berkeley National Laboratory, 1 Cyclotron Road, Berkeley, CA 94720, USA}
    \email{jdbanovetz@lbl.gov}

\author[0000-0003-0691-6688]{Yu-Han Yang}
    \affiliation{Dipartimento di Fisica, Universit\`a di Tor Vergata, Via della Ricerca Scientifica, 1, 00133 Rome, Italy}
    \email{yyang@roma2.infn.it}
    
\author[0000-0002-1869-7817]{Eleonora Troja}
    \affiliation{Dipartimento di Fisica, Universit\`a di Tor Vergata, Via della Ricerca Scientifica, 1, 00133 Rome, Italy}
    \email{eleonora.troja@uniroma2.it}

\author[0000-0002-8680-8718]{Hendrik van Eerten}
\affiliation{Department of Physics, University of Bath, Building 3 West, Bath BA2 7AY, United Kingdom}
\email{hjve20@bath.ac.uk}

\author[0009-0008-2754-1946]{Julius Gassert}
    \affiliation{University Observatory, Faculty of Physics, Ludwig-Maximilians-Universität München, Scheinerstr. 1, 81679 Munich, Germany}
    \affiliation{McWilliams Center for Cosmology and Astrophysics, Department of Physics, Carnegie Mellon University, Pittsburgh, PA 15213, USA}
    \email{julius.gassert@campus.lmu.de}

\author[0009-0001-5050-6232]{Mitra Maleki}
    \affiliation{University Observatory, Faculty of Physics, Ludwig-Maximilians-Universität München, Scheinerstr. 1, 81679 Munich, Germany}
    \email{Mitra.Maleki@campus.lmu.de}
    

\author[0000-0003-4162-6619]{Stephen Bailey}
    \affiliation{Lawrence Berkeley National Laboratory, 1 Cyclotron Road, Berkeley, CA 94720, USA}
    \email{stephenbailey@lbl.gov}

\author[0000-0001-5537-4710]{Segev BenZvi}
    \affiliation{Department of Physics \& Astronomy, University of Rochester, 206 Bausch and Lomb Hall, P.O. Box 270171, Rochester, NY 14627-0171, USA}
    \email{sbenzvi@ur.rochester.edu}

\author[0000-0002-1270-7666]{Tom\'as Cabrera}
    \affiliation{McWilliams Center for Cosmology and Astrophysics, Department of Physics, Carnegie Mellon University, Pittsburgh, PA 15213, USA}
    \email{tcabrera@andrew.cmu.edu}

\author[0009-0000-4830-1484]{Keerthi Kunnumkai}
    \affiliation{McWilliams Center for Cosmology and Astrophysics, Department of Physics, Carnegie Mellon University, Pittsburgh, PA 15213, USA}
    \email{kkunnumk@andrew.cmu.edu}

\author{Adam D. Myers}
    \affiliation{Department of Physics \& Astronomy, University  of Wyoming, 1000 E. University, Dept. 3905, Laramie, WY 82071, USA}
    \email{amyers14@uwyo.edu}

\author[0009-0006-6697-8548]{Christoph Ries}
    \affiliation{University Observatory, Faculty of Physics, Ludwig-Maximilians-Universität München, Scheinerstr. 1, 81679 Munich, Germany}
    \email{cries@usm.lmu.de}

\author[0000-0002-5042-5088]{David Schlegel}
    \affiliation{Lawrence Berkeley National Laboratory, 1 Cyclotron Road, Berkeley, CA 94720, USA}
    \email{djschlegel@lbl.gov}

\author[0009-0003-1323-9774]{Michael Schmidt}
    \affiliation{University Observatory, Faculty of Physics, Ludwig-Maximilians-Universität München, Scheinerstr. 1, 81679 Munich, Germany}
    \email{mschmidt@usm.lmu.de}

\author{Silona Wilke}
    \affiliation{University Observatory, Faculty of Physics, Ludwig-Maximilians-Universität München, Scheinerstr. 1, 81679 Munich, Germany}
    \email{Silona.Wilke@lmu.de}

\author[0009-0004-9520-5822]{Muskan Yadav}
    \affiliation{Dipartimento di Fisica, Universit\`a di Tor Vergata, Via della Ricerca Scientifica, 1, 00133 Rome, Italy} 
    \email{muskan.yadav@students.uniroma2.eu}

\begin{abstract}

The association of broad-lined Type Ic supernovae with long-duration gamma-ray bursts (GRBs) has been known for 28 years.  However, only about seventy gamma-ray burst supernovae (GRB-SNe) have been identified, of which only half have spectroscopic classifications. At $z=0.153$, GRB 260310A is the 12th spectroscopically confirmed GRB-SN discovered within 1 Gpc, offering a critical opportunity to follow one of these rare supernovae in detail. We present optical to near-infrared imaging and spectroscopy of GRB 260310A and SN 2026fgk out to 65 d after discovery. The optical afterglow is among the brightest ever observed from a GRB. Spectra obtained more than two weeks after the explosion reveal broad absorption features that securely identify SN 2026fgk as a Type Ic-BL supernova. Modeling of the multi-wavelength ($grizJK_s$) lightcurve shows that the supernova is approximately half the luminosity ($k_\textrm{98bw}=0.4-0.6$) of the canonical GRB-SN 1998bw. We derive a nickel mass of $M_\textrm{Ni}=0.4-0.5$ $M_\odot$ with a total ejected mass of $M_\textrm{ej}\approx4-6 $ $M_\odot$ and kinetic energy $E_\textrm{K}=(3-8)\times10^{51}$ erg. The GRB exploded at an extremely large offset of 15 kpc from its host galaxy. Long-slit spectra reveal a ``bridge'' of nebular emission extending along the galaxy's disk to the GRB location, which has a sub-solar metallicity ($\sim$\,$0.4Z_\odot$), compared to a near solar metallicity for the host galaxy. This indicates that the large offset arises from the galaxy's extended light profile rather than an isolated environment.

\end{abstract}



\section{Introduction} 
\label{sec:intro}

Long-duration gamma-ray bursts (GRBs) are among the most explosive and energetic objects in the Universe. For the past three decades, these events have been widely understood to mark the deaths of massive stars. The key observational breakthroughs came first from the association of the low-luminosity GRB 980425 to SN 1998bw \citep{Galama1998,Kulkarni1998,Iwamoto1998,Galama1999,Stathakis2000,Patat2001,Kouveliotou2004}, and later from the spectroscopic identification of SN 2003dh \citep{Stanek2003,Hjorth2003,Matheson2003,Mazzali2003,Deng2005} in the afterglow of the more energetic, cosmological burst GRB 030329 \citep{Berger2003,Sheth2003,Vanderspek2004,Taylor2004,Lipkin2004,vanderhorst2008}. These events established that the supernovae accompanying long GRBs (hereafter, GRB-SN) are typically broad-lined Type Ic (hereafter, Type Ic-BL) explosions. These are stripped-envelope core-collapse supernovae whose spectra show very broad absorption features (typically Fe\,{\sc ii} and Si\,{\sc ii}), implying high ejecta velocities ($>10,000$ km s$^{-1}$) and large kinetic energies \citep{Woosley2006,Hjorth2012sn,Hjorth2013,Cano2017}. Over the past two decades our understanding of GRB-SN has evolved to include a broader diversity of supernova properties, though they tend to a narrow range of luminosities and ejecta properties \citep[e.g.,][]{Cano2017} that do not appear to correlate with the GRB jet's energetics \citep[e.g,][]{Hjorth2013}. To date, there are roughly 70 GRB-SN associations. Roughly half of these GRB-SN have spectroscopic classifications \citep[see, e.g.,][and references therein]{Hjorth2012sn,Cano2017,GRBSNtool}, while the remaining candidates are based on late-time photometric bumps in their optical lightcurves \citep[e.g.,][]{Bloom1999-980326,Zeh2004}.

The association of GRBs to stripped-envelope supernovae fits within the collapsar framework for long GRB production \citep{Woosley1993,MacFadyen1999,MacFadyen2001}. In this model, the core collapse of a rapidly rotating, stripped star powers both a relativistic jet and a supernova. Material falling onto a newly formed black hole, or in some cases a rapidly spinning magnetized compact remnant, launches a jet that drills through the star and produces the prompt gamma-ray emission \citep[see, e.g.,][]{Piran2004}, while radioactive heating from freshly synthesized $^{56}$Ni powers the optical supernova \citep[see, e.g.,][]{Woosley2006,Hjorth2012sn,Cano2017}. The observational association between long GRBs and supernovae is therefore a cornerstone of our understanding of jet-driven stellar death.

Low redshift events are especially valuable because they provide the clearest opportunity to disentangle afterglow, host galaxy, and supernova light, and to follow the spectroscopic evolution in detail. GRB 260310A is the latest nearby burst to offer this opportunity. The event was detected by \textit{Fermi} \citep{GCN43951,2026GCN.43975....1H} and \textit{AstroSat} \citep{2026GCN.43958....1S}, although the initial localization was poor. An optical counterpart was discovered by GOTO \citep{GOTOdisco} and reported to the Transient Name Server (TNS)\footnote{\url{https://www.wis-tns.org/object/2026fgk}}, and subsequent optical follow up strengthened its association with GRB 260310A \citep{HindsDisco,GCN1}. The identification was then secured by the discovery of X-ray and radio counterparts \citep{X1,R1,R2,Christy2026}, which established that the optical transient was indeed the afterglow of the burst.

A redshift of $z$\,$=$\,$0.153$ was measured from nebular emission lines \citep{GCN43977,2026GCN.43984....1D,2026GCN.43986....1H}. At this distance, the burst had an isotropic equivalent gamma-ray energy release of $E_{\gamma,\textrm{iso}}$\,$=$\,$3.5\times10^{50}$ erg and a rest frame spectral peak energy of $E_\textrm{p}$\,$=$\,$200$ keV \citep{GCN43951,GCN43977,GCN43981}, placing it as a significant outlier to the Amati relation \citep{Amati2002,Amati2006}. Owing to its proximity, the optical afterglow is one of the brightest ever observed from a GRB \citep{GCN8}. At later times, optical spectroscopy revealed the emergence of a Type Ic-BL supernova, SN 2026fgk, in the light curve \citep{2026GCN.44124....1D,2026GCN.44137....1O,Gill2026,Hinds2026}, conclusively linking GRB 260310A to the collapse of a massive stripped star. GRB 260310A/SN 2026fgk is therefore only the 12th GRB-SN discovered within 1 Gpc. In this paper we present a photometric and spectroscopic characterization of the optical to near-infrared afterglow of GRB 260310A and of the associated supernova SN 2026fgk, place the event in the context of the known GRB-SN population, and use it to extend the still limited sample of nearby, spectroscopically secure GRB-SN.

Throughout the manuscript we adopt a standard $\Lambda$CDM cosmology \citep{Planck2020} with $H_0$\,$=$\,$67.4$ km s$^{-1}$ Mpc$^{-1}$, $\Omega_\textrm{m}$\,$=$\,$0.315$, and $\Omega_\Lambda$\,$=$\,$0.685$. At redshift $z$\,$=$\,$0.153$, this cosmology corresponds to a luminosity distance of 756 Mpc. All upper limits are reported at the $3\sigma$ level and all magnitudes are in the AB system.

\section{Observations}
\label{sec:obs}

\subsection{Fraunhofer Telescope Wendelstein (FTW)}
\label{sec:FTW}

We observed the optical and near-infrared (OIR) counterpart of GRB 260310A with the Three Channel Imager (3KK; \citealt{2016SPIE.9908E..44L}) mounted on the 2.1-m Fraunhofer Telescope at Wendelstein Observatory (FTW; \citealt{2014SPIE.9145E..2DH}). Observations began on 2026-03-13, corresponding to 3.6 d after the trigger. The 3KK camera is capable of obtaining images in three filters simultaneously: two optical and one near-infrared (NIR). We carried out daily observations in the $griz$ bands as weather allowed. NIR observations ($JK_s$) were initially obtained simultaneously, but were not available after 2026-03-28 due to an issue with the cooling system. The cooling system was repaired and NIR observations began again on 2026-04-22. The complete log of observations is tabulated in Table \ref{tab:grb260310a_photometry}. 

The FTW data were reduced using a custom pipeline \citep{2002A&A...381.1095G} to perform bias and dark subtraction, flat-fielding, and cosmic ray rejection. For more details, specifically with regard to the near-infrared data reduction, see \citet{Busmann2025}. For optical images, we performed difference imaging with the Saccadic Fast Fourier Transform (\texttt{SFFT}) software\footnote{\url{https://github.com/thomasvrussell/sfft}} \citep{Hu2022} using archival PS1 \citep{Chambers2016} and Legacy Survey images as templates.  Due to the large host offset, the non-difference imaging photometry was nearly identical ($<0.1$ mag deviations) to difference imaging photometry until late times ($\gtrsim$15 d). Due to the lack of available NIR templates, the $JK_s$ photometry did not undergo difference imaging. We performed aperture photometry on the difference images using \texttt{Photutils} \citep{Bradley2024} with AB magnitude zeropoints calibrated to the PS1 \citep{Chambers2016} and 2MASS \citep{Skrutskie2006} catalogs for optical and near-infrared data, respectively. Standard Vega to AB magnitude conversions were applied to the NIR data. The photometry is reported in Table \ref{tab:grb260310a_photometry}.

\subsection{Seimei Telescope}

We obtained low-resolution spectra at two epochs using the Kyoto Okayama Optical Low-dispersion Spectrograph with optical-fiber Integral Field Unit (KOOLS-IFU; \citealt{2019PASJ...71..102M,2025PASJ...77.1065M}), mounted on the 3.8-m Seimei telescope on Mt. Chikurinji, Japan. We used VPH-blue (2026-03-14) and VPH-red (2026-03-15) grisms, giving a spectral resolution of $\lambda / \Delta \lambda \simeq$ 500 and 700, respectively. After a standard reduction procedure including bias subtraction, flat-field correction, distortion correction, and cosmic-ray removal, a one-dimensional spectrum was extracted from the two-dimensional image by subtracting a sky spectrum extracted from the region around the target. For the wavelength calibration, we used Hg, Ne, and Xe lamp frames. Finally, we performed flux calibration by observing a spectrophotometric standard star (HR3454). A log of spectral observations is presented in Table \ref{tab:grb260310a_spectra}.

We obtained $g$-, $r$-, $i$-, and $z$-band images using the TriColor CMOS Camera and Spectrograph (TriCCS) mounted on the 3.8-m Seimei telescope starting on 2026-03-14. After standard image reduction procedures including bias, dark and flat corrections, and astrometric alignment, the multiple images in each filter taken on the same night are stacked based on the astrometric solution. Difference imaging and photometry was performed using the same pipeline and methods as outlined above for FTW (see \S \ref{sec:FTW}). The photometry is reported in Table \ref{tab:grb260310a_photometry}.

\subsection{DESI}

The Dark Energy Spectroscopic Instrument (DESI) includes of 5,000 science fibers that can each be positioned independently on sky \citep{schlafly_survey_2023,poppett_overview_2024}. Each fiber is 107 $\mu$m in diameter, which is equivalent to $\sim1.5\arcsec$ on the sky. The spectrograph utilizes three cameras in the approximate $BRZ$ bands that cover $3600-9824$ \AA\ at a spectral resolution $R \sim 2000 - 5500$ \citep{desi_collaboration_desi_2016, miller_optical_2024, desi_collaboration_data_2025, desi_collaboration_desi_2025}. Each spectrum is reduced and flux calibrated with the DESI spectroscopic data pipeline \citep{guy_spectroscopic_2023}. 

DESI obtained a sequence of spectra of GRB 260310A (see Table \ref{tab:grb260310a_spectra}) at 6.2, 7.1, 8.1, and 12.1 days after the GRB discovery. Each spectrum had an effective DESI exposure of 1200 s as part of a ``dark'' time tile. The spectra were obtained as part of the DESI Transients Survey \citep{HallDTS}, a spare fiber program that targets publicly reported transients. For other examples of DESI spare fiber targets and a more in depth review of the observing strategy, see \citet{Myers:2022azg,DESI:2023ytc,HallDTS,HallDESIulz}. The DESI spectra are publicly available through the DESI Transients Survey Zenodo \citep{hall_2026_19653826}. The DESI spectra are automatically corrected for Galactic extinction, which is minimal in this field $E(B-V)=0.02$ mag \citep{Schlafly2011}.

\begin{figure}
    \centering
    \includegraphics[width=\linewidth]{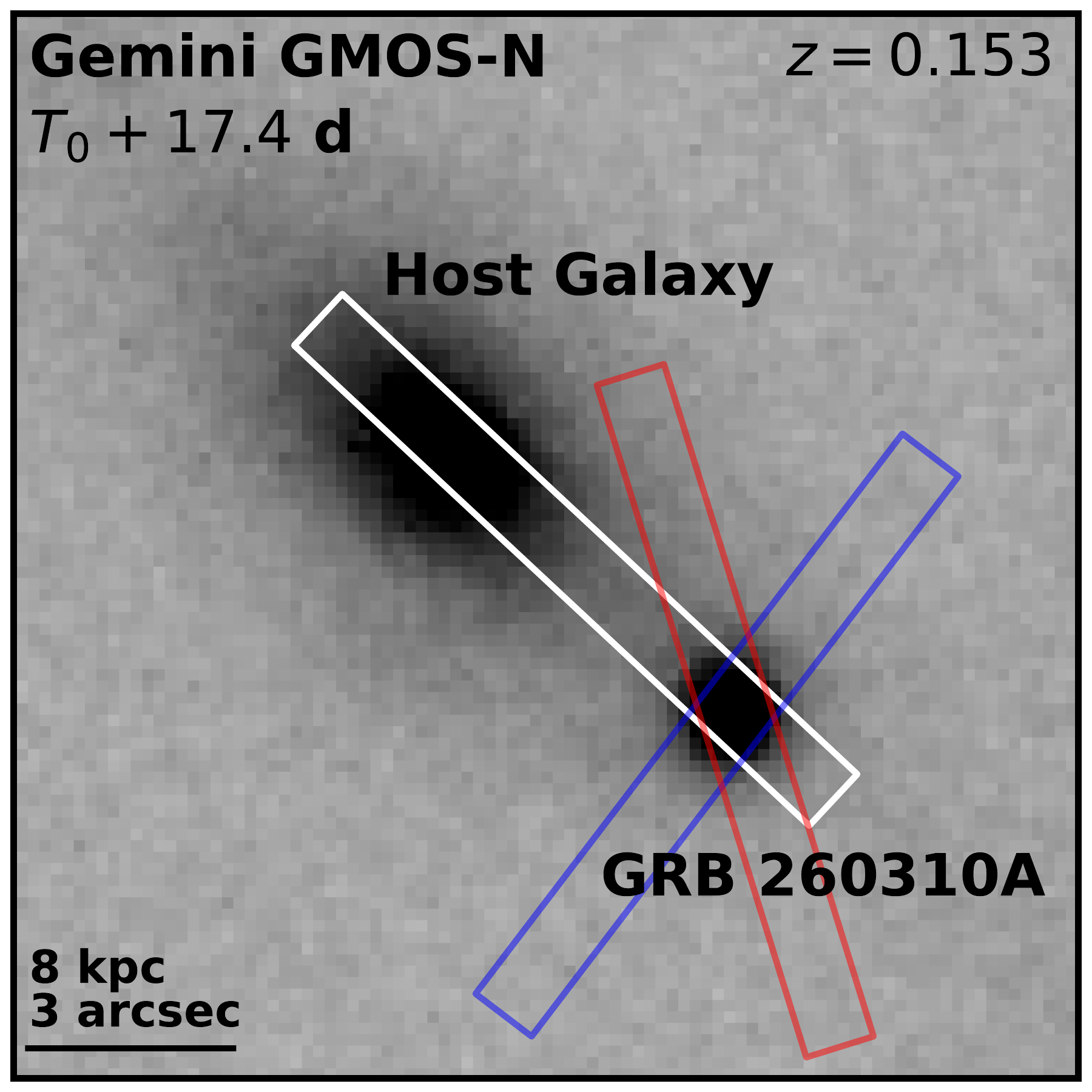}
    \caption{Gemini $r$-band acquisition image ($3\times15$ s) from 2026-03-27 ($T_0+17.4$ d) showing the placement of the Gemini spectroscopic long-slit (white) observation that covers both the GRB and the host galaxy. Later Gemini spectra were obtained using different position angles (blue for 2026-04-23; red for 2026-05-13). The length of the rectangles mimicking the slit are simply for visualization purposes, while the $1\arcsec$ width is shown to scale. 
    North is up and East is to the left.
    }
    \label{fig:GemFC}
\end{figure}

\subsection{HET}

We observed GRB 260310A with the 11-m Hobby-Eberly Telescope (HET; \citealt{1998SPIE.3352...34R, 2021AJ....162..298H}) at McDonald Observatory under program M26-1-005 (PI: D. Gruen) on 2026-03-19). The observations were scheduled using the HET queue-scheduling system \citep{2007PASP..119..556S}. We used the  low-resolution integral-field spectrograph (LRS2; \citealt{Chonis2014,Chonis2016}) to obtain spectra in the red channel LRS2-R data. 
The raw LRS2 data were initially processed with \texttt{Panacea}\footnote{\url{https://github.com/grzeimann/Panacea}}, which performs bias subtraction, dark subtraction, fiber tracing, fiber wavelength evaluation, fiber extraction, fiber-to-fiber normalization, source detection, source extraction, and flux calibration for each channel. The absolute flux calibration was based on default response curves, measurements of mirror illumination, and exposure throughput derived from guider images. We extracted the flux-calibrated one-dimensional spectrum using the \texttt{LRS2Multi}\footnote{\url{https://github.com/grzeimann/LRS2Multi}} package, integrating over the $0.59\arcsec$ fibers within a $1.5\arcsec$ radius aperture centered on the transient (Figure \ref{fig:GemFC}). 

\subsection{Gemini}

We performed observations with the Gemini Multi-Object Spectrographs (GMOS) at Gemini North under programs GN-2026A-Q-202 (PI: M. Tanaka), GN-2026A-DD-102 (PI: B. O'Connor), and GN-2026A-Q-123 (PI: G. Srinivasaragavan). The observations were performed under program GN-2026A-DD-102 (PI: B. O'Connor) with the time charged across multiple programs. Longslit spectroscopy of the optical transient was acquired on 2026-03-27 (17.4 d; PIs: Tanaka and O'Connor), 2026-04-05 (26.3 d) and 2026-04-20 (41.3 d), 2026-04-23 (44.3 d), and 2026-05-13 (64.1 d), see also Table \ref{tab:grb260310a_spectra}. Due to shared data access of the late spectra, these spectra also appear in \cite{Hinds2026}. All spectra used the B480 grating and a $1\arcsec$ slit width. For the spectra taken on 2026-03-27, 2026-04-05, and 2026-04-20, the slit covered both the transient and the host galaxy 
as shown in Figure \ref{fig:GemFC}. These spectra suffer from slit losses caused by atmospheric differential refraction \citep{Filippenko1982} and have untrustworthy flux calibrations at the blue end, which we exclude from our analysis. The spectra have been absolute flux calibrated based on near-simultaneous photometry, which agrees with the observed spectral shapes at $>$\,$5000$ \AA. Additional corrections for slit losses were made following \citep{Filippenko1982}. Atmospheric differential refraction does not impact the identification of supernova features at $>$\,$5000$ \AA. The slit orientation of later spectra covered only the transient (2026-04-23), and edges of the host galaxy light profile (2026-05-13), and were less impacted by atmospheric differential refraction. The sequences on 2026-03-27 and 2026-05-13 were only partially obtained due changing weather conditions. The data were reduced and analyzed using the \texttt{DRAGONS} software \citep{Labrie2019}.

\section{Results and Analysis}

\subsection{Multi-wavelength Lightcurve Modeling}
\label{sec:redback}

\subsubsection{Model Setup}

We performed a simultaneous multi-wavelength optical to near-infrared (OIR) fit to the lightcurve of GRB 260310A in the $grizJK_s$ filters using models supplied in the \texttt{redback} software package \citep{redback}. Posterior sampling was performed with \texttt{Bilby} \citep{bilby} using the nested sampler \texttt{dynesty} \citep{dynesty}. We empirically modeled the lightcurve with the combination of a afterglow component plus the contribution of a supernova. 

For the supernova we applied two models. The first was a \texttt{redback} model based on the SN 1998bw \citep{Galama1998,Stathakis2000,Patat2001,Sollerman2000,Sollerman2002,Clocchiatti2011}, with an  extrapolation to extend the wavelength range of the SN's blackbody spectrum to redder wavelengths to allow joint fitting of the near-infrared ($JK_s$) data. Due to this model extrapolation, the $JK_s$ model is less certain and prone to larger uncertainties. The \texttt{redback} model is based on the
\texttt{sncosmo} \citep{sncosmo} \texttt{v19-1998bw} template \citep{Vincenzi2019}. This model is parametrized by a simple multiplicative scale factor $k_\textrm{98bw}$ to correct the flux of the template to match observations. A value of $k_\textrm{98bw}$\,$=$\,$1$ matches the observed flux scale of SN 1998bw with rest-frame absolute magnitudes $M_B$\,$=$\,$-18.9$ and $M_V$\,$=$\,$-19.3$ mag \citep{Galama1998}. A second fit instead employed a simple Arnett model \citep{Arnett1982}. This model is parameterized by the nickel fraction $f_{\rm Ni}$, ejecta mass $M_{\rm ej}$, ejecta velocity $v_{\rm ej}$, and floor temperature $T_{\rm floor}$. In both cases, the models were integrated through the $grizJK_s$ filters using \texttt{sncosmo} \citep{sncosmo} to ensure proper K-corrections were employed to match the observed wavelengths at redshift $z$\,$=$\,$0.153$.

The afterglow component was phenomenologically parametrized as a simple temporal broken powerlaw of the form \citep{Beuermann1999,Rhoads2001}:
\begin{equation}
F_{\nu,{\rm AG}}\propto \nu^{-\beta}
\left[
\frac{1}{2}
\left(
\left(\frac{t}{t_{\rm break}}\right)^{s\alpha_1}
+
\left(\frac{t}{t_{\rm break}}\right)^{s\alpha_2}
\right)
\right]^{-1/s},
\end{equation}
where $\beta$ is the spectral slope of the afterglow, $s$ is the smoothing parameter, and $\alpha_1$ and $\alpha_2$ are the temporal slopes before and after the break time $t_{\rm break}$. We jointly fit the X-ray lightcurve out to 18.5 d \citep{Yang2026} in order to constrain the spectral index and post-break temporal slope, excluding the rebrightening phase (see \S \ref{sec:XOIR-compare} for further discussion).

We performed both sets of fits (powerlaw+SN 1998bw and powerlaw+Arnett), which are displayed in Figure \ref{fig:AGSNfit} and Figure \ref{fig:AGSNfit-2} in Appendix \ref{sec:modelfitappendix}. Both fits were performed with an additional 10\% systematic error added in quadrature to the statistical error of each datapoint to account for uncertainties (e.g., different filter transmission curves) between the different telescopes used.

\begin{figure}
    \centering
    \includegraphics[width=\linewidth]{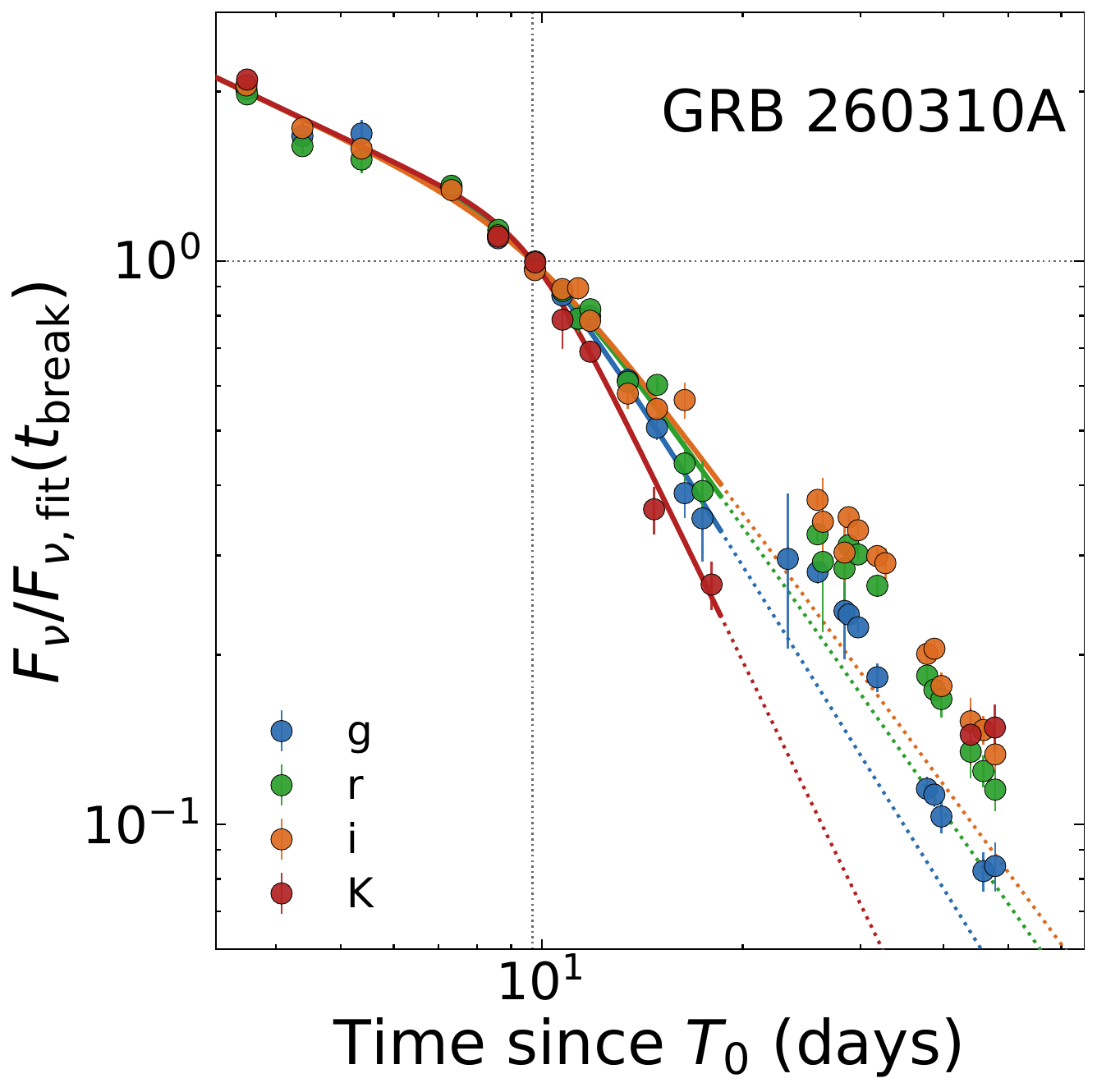}
    \caption{Normalized flux density relative to the flux density at the break time $t_\textrm{break}$ for the $griK$ bands. This shows the different relative slopes of these bands after the break, as well as the late-time color evolution.
    }
    \label{fig:normflux}
\end{figure}

\begin{figure*}
    \centering
    \includegraphics[width=\linewidth]{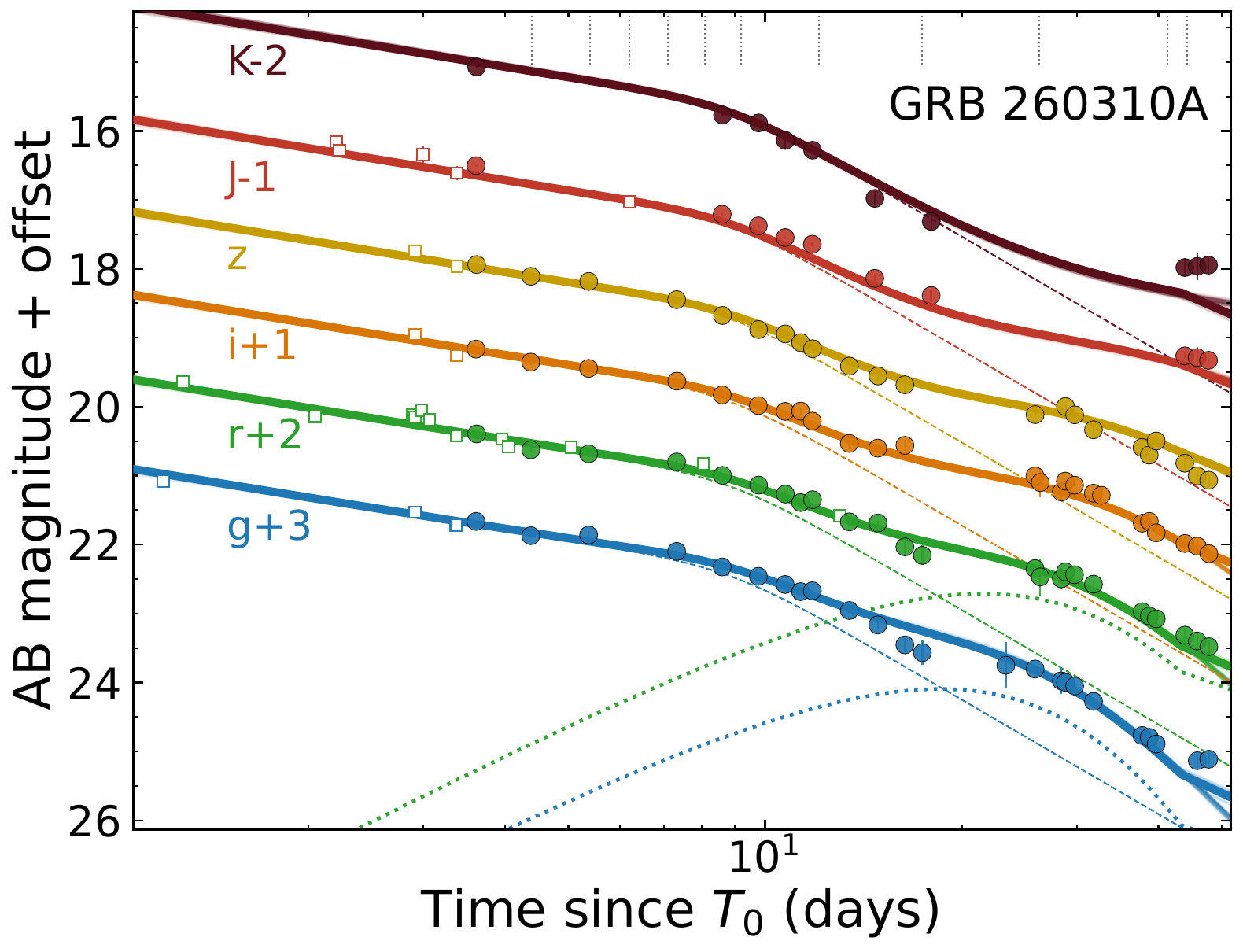}
    \caption{Best-fit model (powerlaw+Arnett) to the multi-wavelength lightcurve of GRB 260310A in the observer-frame. The thick solid lines in each filter represent the total model. The thin dashed lines represent the broken power-law afterglow component and the dotted lines, shown only in the $g$- and $r$-bands to avoid clutter, represents the Arnett model lightcurve. Photometry presented in this work (FTW and Seimei) is shown as a solid circle and publicly reported photometry compiled from GCN Circulars is shown as empty squares \citep{HindsDisco,GCN1,GCN2,GCN3,GCN4,GCN5,GCN6,GCN7,GCN8,GCN9,GCN10,GCN11,GCN12}. The vertical lines at the top of the figure mark the epochs of our optical spectra (Figure \ref{fig:specsequence}). 
    }
    \label{fig:AGSNfit}
\end{figure*}

\subsubsection{Comparison of X-ray and OIR Behavior}
\label{sec:XOIR-compare}

The OIR lightcurve is well described by an initial shallowly decaying powerlaw ($\approx$\,$t^{-0.65\pm0.07}$) with a temporal break to a steeper decay at $\sim$\,$8-10$ d (observer frame) after the GRB trigger. The post-break $K_s$ lightcurve shows a steeper decay ($\approx$\,$t^{-2.00\pm0.15}$) than the optical data ($\approx$\,$t^{-1.65\pm0.10}$ in $g$ and $\approx$\,$t^{-1.47\pm0.10}$ in $z$ and $J$). We show this in Figure \ref{fig:normflux}, where the flux density in the $griK$ bands is shown normalized to the flux at the break. This shows clearly the different post-break slopes in each band.

A similar behavior is also observed at X-rays \citep{xrayslopes1}, where an initial shallow slope $\approx$\,$t^{-0.35\pm0.08}$ breaks to a steeper decay $\approx$\,$t^{-2.05\pm0.15}$ after $\sim$\,$8-10$ d \citep{Yang2026}. Thus, we find that the post-break X-ray decay slope matches the $K_s$-band slope. The different post-break decay slopes of the OIR bands can be explained by the contribution of a supernova to the optical lightcurve, which serves to flatten out the powerlaw decay prior to taking over nearly completely at late times. Instead, the $K_s$-band matches the X-ray decline as the supernova contributes least to the NIR at that stage of its evolution. This signals that the pre-break lightcurve is dominated instead by the afterglow, as expected. 
While it is difficult to reconcile the initially shallow decay slopes with standard afterglow closure relations \citep{Sari1998,Wijers1999}, the achromatic break can be interpreted as a jet-break \citep{Rhoads1999,SariPiranHalpern1999,MeszarosRees1999,granot2002}.

Following the achromatic break observed in the X-ray and OIR data, the X-ray lightcurve subsequently exhibited an extremely late-phase plateau (or a rebrightening; \citealt[][]{rebright1,rebright2,Yang2026,Gill2026,Hinds2026}). The exact origin of the late-time X-ray emission is unclear, though it could be potentially explained by an off-axis jet \citep{Rossi02,Granot2002jet,Panaitescu2003,Kumar2003}, a two-component jet \citep{Pedersen1998,Berger2003,Sheth2003,Peng2005}, energy injection \citep{Zhang2001,Zhang2006,Liang2006}, refreshed shocks \citep{ReesMeszaros1998,SariMeszaros2000,KumarPanaitescu2000}, inverse Compton (IC) scattering of optical supernova photons \citep{Chevalier1982,Bjornsson2004,Chevalier2006,Soderberg2012,Margutti2012}, or a long-lived magnetar central engine \citep{Troja2007,Lyons2010,Rowlinson2010}. We note that additional physics beyond the standard forward shock afterglow are also potentially required to match the shallow X-ray and OIR slopes at early times \citep[see][]{Christy2026,Yang2026,Gill2026}. 

As the nature of this additional X-ray emission is not certain, it is difficult to predict its impact on the OIR lightcurve, as depending on the interpretation, this rebrightening may or may not contribute to the OIR emission (e.g., IC SN emission would not contribute to the OIR) at late-times and with varying contributions. We leave a full analysis of the multi-wavelength dataset to future work (see also \citealt{Yang2026}). 

While the timescales of the X-ray rebrightening matches the typical timescales of a Ic-BL supernova, it is difficult to produce such luminous late-time X-ray emission with IC scattering alone, see, e.g., \citet{Margutti2013-100316D}. If instead the rebrightening impacts the entire broadband afterglow, it could contribute significantly to the OIR depending on the spectral index and spectral energy distribution (e.g., a spectral break between the non-thermal X-ray and OIR emission at late-times). We find that while the non-thermal rebrightening is capable of matching the overall OIR brightness, it cannot explain either the \textit{i}) supernova spectral features out to $40$\,$-$\,$60$ d or \textit{ii}) the long-term OIR color evolution (in particular relative to $g$-band; see Figure \ref{fig:normflux}). These features strongly suggest a non-thermal rebrightening cannot be the dominant contribution to the observed OIR emission after 20 d. If the rebrightening does contribute at OIR wavelengths then our inferred supernova contribution should be taken as an upper limit, though the existence of supernova spectral features (see \S \ref{sec:specfeat}) does require a significant supernova component ($k_\textrm{98bw}$\,$\gtrsim$\,$0.2$) to exist between $20$\,$-$\,$60$ d.

If instead the OIR rebrightening at 20+ d is due predominantly to the same component as the X-ray rebrightening, and that the OIR and X-ray lie on the same spectral index, this would imply that SN 2026fgk is potentially the faintest GRB-SN observed to date. While this is possible, it is inconsistent with the spectral features of the SN at 40+ d, and, according to Occam's razor, it is unlikely to have the faintest GRB-SN also appear in a source exhibiting an extremely odd X-ray rebrightening.

\subsubsection{Results of Model Fitting}

Our model fits (powerlaw+SN 1998bw and powerlaw+Arnett) are shown in Figure \ref{fig:AGSNfit} and Figure \ref{fig:AGSNfit-2} in Appendix \ref{sec:modelfitappendix}. The greater flexibility of the simplified Arnett model provides a better match to the supernova evolution (Figure \ref{fig:AGSNfit}). While SN 1998bw provides a useful representative case (Figure \ref{fig:AGSNfit-2} in Appendix \ref{sec:modelfitappendix}), the supernova evolution of SN 2026fgk is clearly at least slightly different, particularly in the redder bands where it is brighter than SN1998bw. Additionally, based on our fit, which yields $k_\textrm{98bw}$\,$\approx$\,$0.4$\,$-$\,$0.6$, we find that SN 2026fgk is approximately $0.5$\,$-$\,$1.0$ mag fainter than SN 1998bw. 

In either case, the exact SN contribution is sensitive to the underlying afterglow contribution (e.g., the late-time temporal index). However, the afterglow decay inferred here agrees well with the X-ray lightcurve out to 18.5 d \citep{Yang2026}. 
Additionally, our inferred supernova ejecta mass of $4$\,$-$\,$6$\,$M_\odot$ with nickel mass $\sim$\,$0.4$\,$-$\,$0.5$\,$M_\odot$, ejecta velocity of $\sim$\,$16,000\pm3,000$ km s$^{-1}$, and total kinetic energy $E_\textrm{k}$\,$=$\,$(3-8)\times10^{51}$ erg are in good agreement with other GRB-SN observations \citep{Hjorth2012sn,Cano2013,Cano2017}, see, e.g., Fig. 5 of \citet{Srinivasaragavan2023}.

\begin{figure*}
    \centering
    \includegraphics[width=\linewidth]{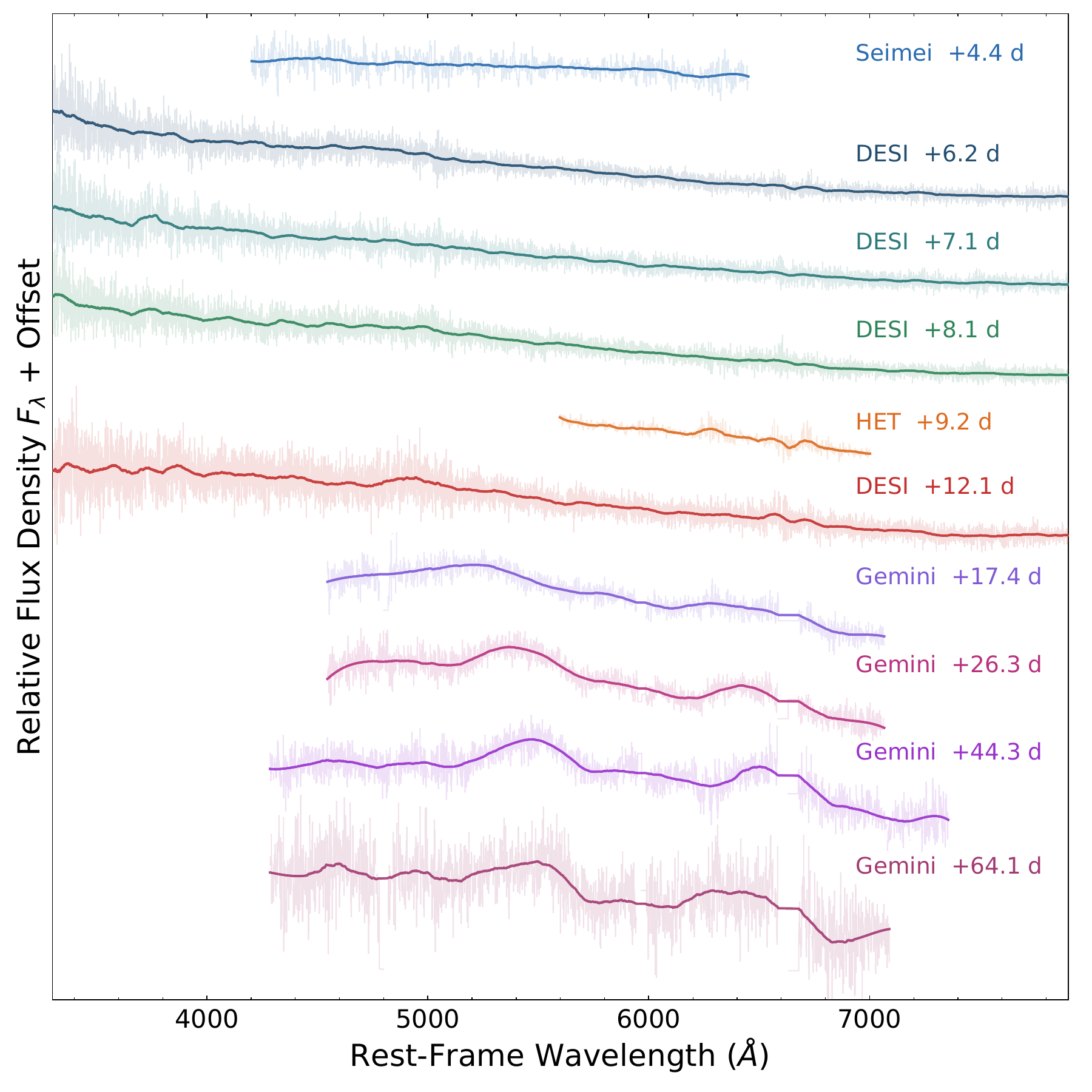}
    \caption{Spectral sequence of GRB 260310A obtained with Seimei, DESI, HET, and Gemini. The spectra are smoothed with a Savitzky-Golay filter (thick lines) and the unsmoothed spectra (thin lines) are also shown. Emission lines and telluric absorption regions 
    have been clipped from the spectra prior to smoothing. The spectra are initially well described by a powerlaw $F_\nu$\,$\propto$\,$\nu^{-\beta}$ with spectral index $\beta$\,$\approx$\,$1.0$ and later evolve to show broad absorption features at $\sim5000\,\AA$ and $\sim6100\,\AA$. 
    }
    \label{fig:specsequence}
\end{figure*}

\begin{figure*}
    \centering
    \includegraphics[width=\linewidth]{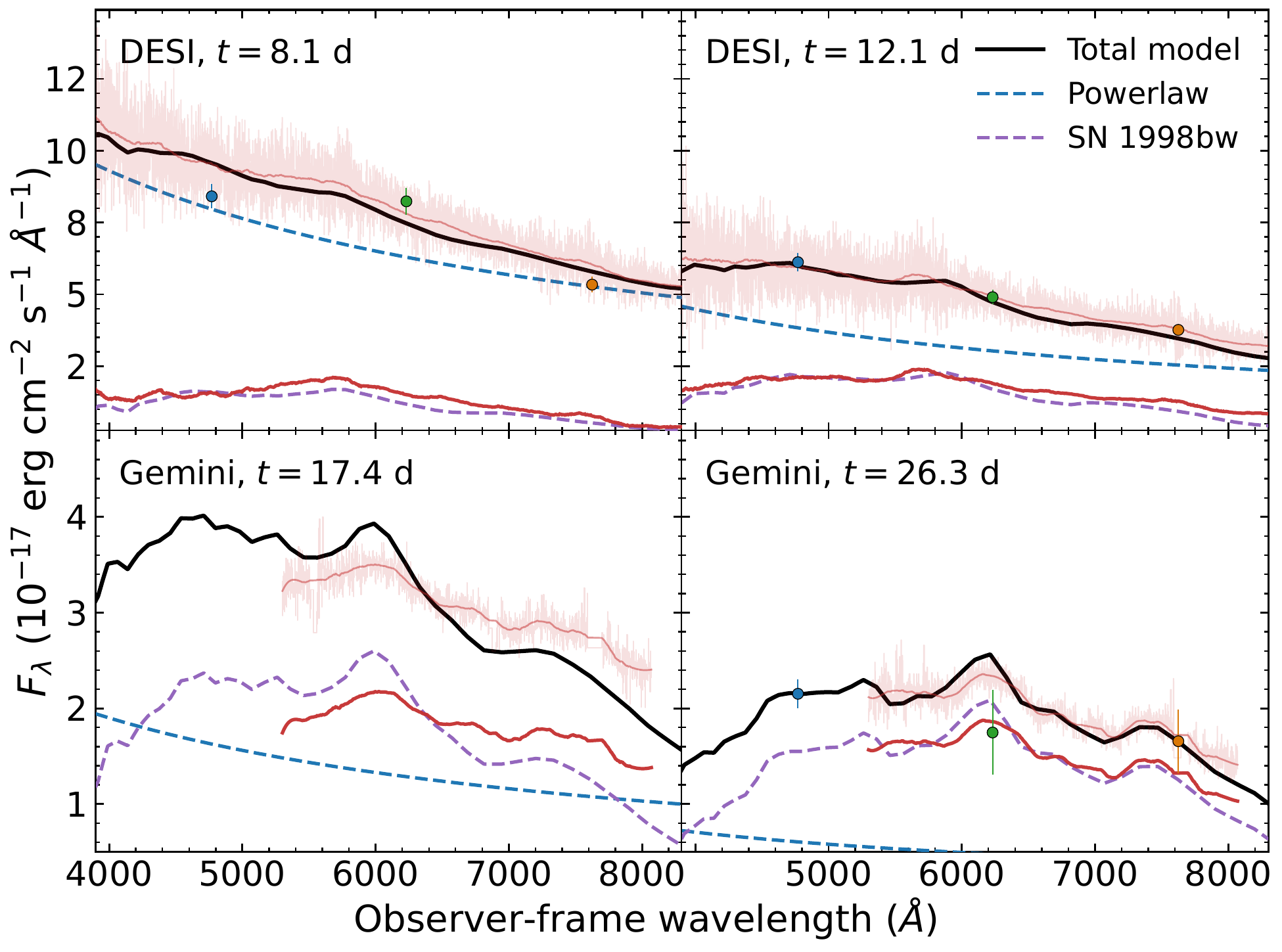}
    \caption{Comparison between the best-fit model derived only from the photometry to the observed DESI and Gemini spectra at 8.1, 12.1, 17.4, and 26.3 d showing the onset and evolution of supernova features. The solid black line represents the total model (powerlaw+SN1998bw) while the blue dashed line represents the powerlaw component ($F_\nu$\,$\propto$\,$\nu^{-\beta}$) and the dashed purple line represents the spectral series of SN 1998bw with a flux scaling factor $k_\textrm{98bw}$\,$=$\,$0.50$. In each panel the observed spectrum is shown both in raw form and smoothed. The thick solid red line represents the smoothed observed spectrum with the powerlaw component subtracted. The marker color reflects the bandpass of the photometry using the same color scheme as in Figure \ref{fig:AGSNfit}.
    }
    \label{fig:modelspeccompare}
\end{figure*}

\subsection{Spectral Analysis and Supernova Features}
\label{sec:specfeat}

We obtained an optical spectral sequence (Figure \ref{fig:specsequence}) of GRB 260310A starting at 4.4 d and continuing to 64 d in the observer frame.  The timing of our spectral sequence relative to the lightcurve is shown in Figure \ref{fig:AGSNfit}. The early spectra from Seimei, DESI, and HET are well described by a powerlaw $F_\nu$\,$\propto$\,$\nu^{-\beta}$ with spectral index $\beta$\,$\approx$\,$1.0$. 
This agrees with the photometry based spectral energy distributions (SEDs) at $<$\,$8$ d and our lightcurve modeling in \S \ref{sec:redback}. This observed spectral index can also be explained by $\beta$\,$\approx$\,$0.9$ extincted by a visual extinction $A_\textrm{V}$\,$\approx$\,$0.1$ mag from dust intrinsic to the source environment at $z$\,$=$\,$0.153$, which provides a better agreement with the X-ray data \citep{Yang2026}. 

\begin{figure*}
    \centering

    \includegraphics[width=\linewidth]{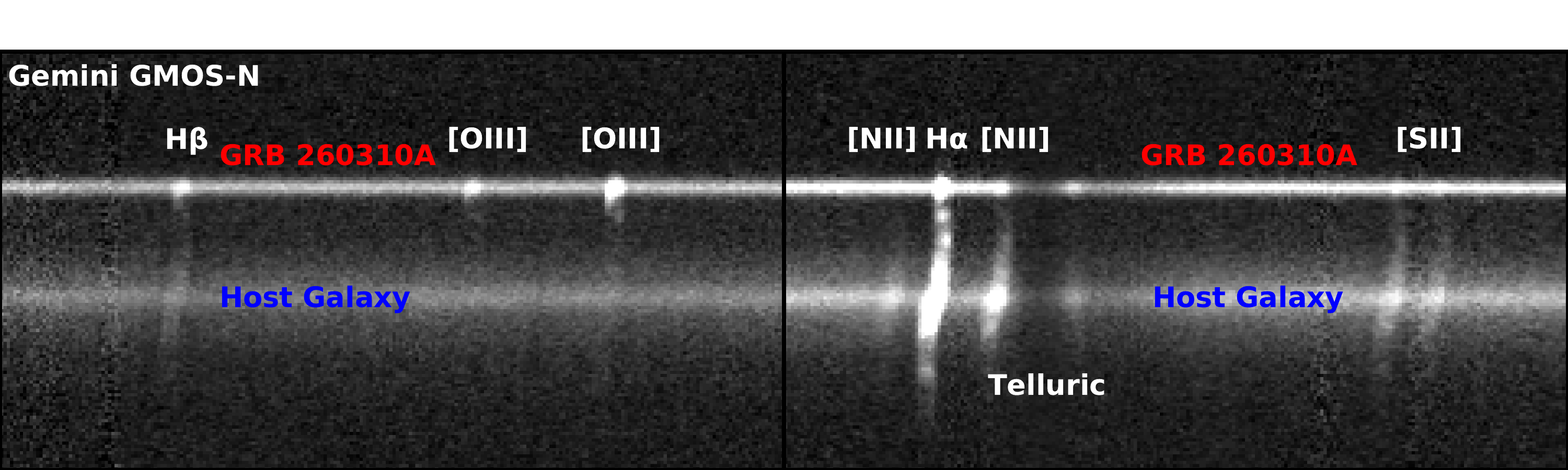}
    \caption{
    Visualization of the 2D Gemini GMOS-N spectrum obtained on 2026-04-05 ($T_0+26.3$ d). The spectrum covers both the host galaxy (bottom trace) and transient (top trace), as shown in Figure \ref{fig:GemFC}. The spectral direction is the X-axis which goes from blue to red wavelengths from left to right. The spatial direction along the slit is the Y-axis. Each panel covers 300 \AA\, (observer frame) in width and is 11\arcsec\, in height. The scale is chosen to maximize the visibility of the emission lines. An atmospheric telluric absorption region (the A-band) is visible to the right of [NII].
    }
    \label{fig:Gem2D}
\end{figure*}

The spectra clearly evolve and display broad undulations typical of Type Ic-BL supernovae at later times. We associate these with blueshifted Fe\,{\sc ii} $\lambda5169$ and Si\,{\sc ii} $\lambda6355$ absorption at high velocities $>10^4$ km s$^{-1}$, typical of GRB-SN. The later spectra obtained with Gemini are noisy and suffer from telluric absorption and chip gaps in critical regions of the Si\,{\sc ii} absorption features. However, the evolution of the features is clear. In fact, we find evidence for deviations from the powerlaw in spectra obtained at 8.1 d and 12.1 d by DESI. These features agree well with the expectations from our powerlaw+SN 1998bw fit to the photometry, which included no additional spectral information. This is shown in Figure \ref{fig:modelspeccompare}, which displays the spectral energy distribution of the powerlaw+SN 1998bw fit to the OIR lightcurves (see \S \ref{sec:redback}) at 8.1, 12.1, 17.4 and 26.3 d versus the relevant DESI and Gemini spectra. We show the total model and the individual components. Additionally, we have subtracted the powerlaw component from the observed spectra after smoothing. This shows good agreement with the features and evolution of SN 1998bw. The spectra suggest that the velocity of the Fe\,{\sc ii} feature was initially faster than SN 1998bw at 8.1 and 12.1 d, but slowed down by 17.4 and 26.3 d to below the speed of SN 1998bw's ejecta. While this is not an exact template match to the observed spectra, it provides a useful diagnostic and comparison to the canonical GRB-SN. This comparison, and the observed spectra, provide conclusive evidence than GRB 260310A was associated with a standard collapsar progenitor that produced a Type Ic-BL GRB-SN. Additionally, the supernova features definitively show that the optical rebrightening has to have a significant, likely dominant, supernova contribution. 

The supernova features continue to be clearly seen out to late times. Figure \ref{fig:specsequence} shows the presence of the bumpy supernova spectra at 44.3 and 64.1 d. We note that the final Gemini spectrum at 64.1 d is noisier due to a combination of fading of the source and that only half the planned sequence was executed due to changing weather conditions. However, despite this, the bump at 5500 \AA\, is still clearly visible. Between 44.3 and 64.1 d the Fe\,{\sc ii} $\lambda5169$ features gradually declined, as observed for other nearby Ic-BL supernovae at similar distances \citep[see, e.g.,][]{Rastinejad2025EP,Srinivasaragavan2025EP0108a,Srinivasaragavan2026}.

\subsection{Environmental Analysis of GRB Site and Galaxy}
\label{sec:host}

GRB 260310A is located at a large angular offset of $5.4\pm0.2\arcsec$ from its host galaxy (Figure \ref{fig:GemFC}). We obtained Gemini spectra (Table \ref{tab:grb260310a_spectra}) that covered both the transient and host galaxy with the slit aligned as shown in Figure \ref{fig:GemFC}. A visualization of the 2D spectra is shown in Figure \ref{fig:Gem2D} that clearly demonstrates that the GRB explosion site and nearby galaxy are located at $z$\,$=$\,$0.1530$ and $z$\,$=$\,$0.1525$ with only a small velocity offset of $\sim$\,$100$ km s$^{-1}$ between them, likely due to the galaxy's rotation. This velocity offset, as well as the rotation curve of the galaxy, are visible in Figure \ref{fig:Gem2D}. At this redshift, the projected physical offset between the galaxy's centroid and the GRB explosion site is $14.8\pm0.6$ kpc. Figure \ref{fig:Gem2D} shows a ``bridge'' of star formation reaching from the galaxy to the GRB site along the slit. The next closest knot of star formation observed along the slit is 1.3 kpc from the GRB's explosion site. Tracing the galaxy rotation curve of H$\alpha$ back along the slit, we find that even in the opposite direction the ongoing star formation in the galaxy's disk extends to at least $6\arcsec$ from the center of the galaxy. This is consistent with the offset of the GRB, and shows that its observed location fits within the light profile of the galaxy, as traced by ongoing star formation. Additionally, deep imaging ($g$\,$>$\,$25$ AB mag) obtained with FTW/3KK at later times clearly shows the morphology of the galaxy is that of a central bulge superimposed on a disk that extends to the GRB explosion site.

We performed an analysis of the nebular emission lines of both the GRB site and host galaxy to explore their individual properties. The GRB explosion site shows common nebular emission lines associated with ongoing star formation (Figure \ref{fig:Gem2D}), including the [O \textsc{ii}] doublet at $\lambda\lambda3726,3729$, nebular [O \textsc{iii}] at $\lambda\lambda4959,5007$, Balmer lines (H$\alpha$,  H$\beta$, H$\gamma$, H$\delta$, H$\epsilon$, and H$\zeta$), [Ne \textsc{iii}] at $\lambda3869$, He I $\lambda5876$, and the [N \textsc{ii}] and [S \textsc{ii}] doublets. The host galaxy shows weaker emission features and deep Balmer and calcium absorption that together indicate both an old stellar population and a new burst of star formation. For the host, H$\beta$ is detected in both emission and absorption with a slight velocity shift in the centroid of those components. This may be due to the rotation curve of the galaxy, but likely also shows that the kinematics (and locations) of the old and new stellar populations in the host galaxy are different. We also detect Na \textsc{i} D absorption from the host galaxy, but not at the location of the transient. 

We used the penalized pixel-fitting (\texttt{pPXF}) software \citep{ppxf,ppxf2,ppxf3} to fit the optical spectra and derive the relevant emission line fluxes. We utilized the \texttt{E-MILES} stellar library \citep{EMILES}. At the transient location, we derive a near Case B recombination Balmer decrement H$\alpha/$H$\beta$\,$=$\,$2.76\pm0.12$ \citep{Osterbrock1989}, Baldwin, Phillips \& Terlevich (BPT; \citealt{BPT}) ratios of $\log\big([$O\,{\sc iii}$]/$H$\beta\big)$\,$=$\,$0.46\pm0.02$ and $N2\equiv\log\big([$N\,{\sc ii}$]/$H$\alpha\big)$\,$=$\,$-0.95 \pm 0.03$ placing the transient location firmly in the star forming region, a subsolar gas phase metallicity 12 + log(O/H) $\approx$\,$8.2$\,$-$\,$8.3$ (roughly $0.4\,Z_\odot$) derived from N2 and O3N2 \citep{PettiniPagel2004,Marino2013}, where we adopt a solar metallicity of 12 + log(O/H) $=$\,$8.69$ \citep{Asplund2004,Asplund2009}. The statistical uncertainty ($\sim$\,$\pm0.05$) on the metallicity is smaller than the systematic uncertainty (0.1-0.2 dex) on the calibration \citep{PettiniPagel2004,Marino2013}. 
From the H$\alpha$ luminosity at the transient location, we derive a star formation rate (SFR) of $0.85\pm0.10\, M_\odot$ yr$^{-1}$ \citep{Kennicutt1998} using a Chabrier \citep{Chabrier2003} initial mass function (IMF).

For the host galaxy, we derived $\log([{\rm N\,II}]/{\rm H}\alpha)$ $=$ $-0.34 \pm 0.01$ and $\log([{\rm O\,III}]/{\rm H}\beta)$ $=$ $-0.44 \pm 0.05$. From these, we obtain a near solar metallicity of $12+\log({\rm O/H}) \approx 8.6$\,$-$\,$8.7$ from O3N2 and N2 \citep{PettiniPagel2004,Marino2013}. Therefore, the host galaxy has a higher metallicity than the GRB explosion site. While these quoted errors do not include the intrinsic calibration scatter, which is of order $\sim$\,$0.18$ dex for O3N2 and $\sim$\,$0.16$ dex for N2 from \citet{Marino2013}, both diagnostics show the same trend that the GRB environment has a lower metallicity by roughly $0.3$\,$-$\,$0.5$ dex. In the context of the comparison sample shown in Figure \ref{fig:bpt}, the local GRB site sits in the region more commonly occupied by resolved GRB explosion sites compared to the integrated host value. 

\begin{figure}
    \centering
    \includegraphics[width=\columnwidth]{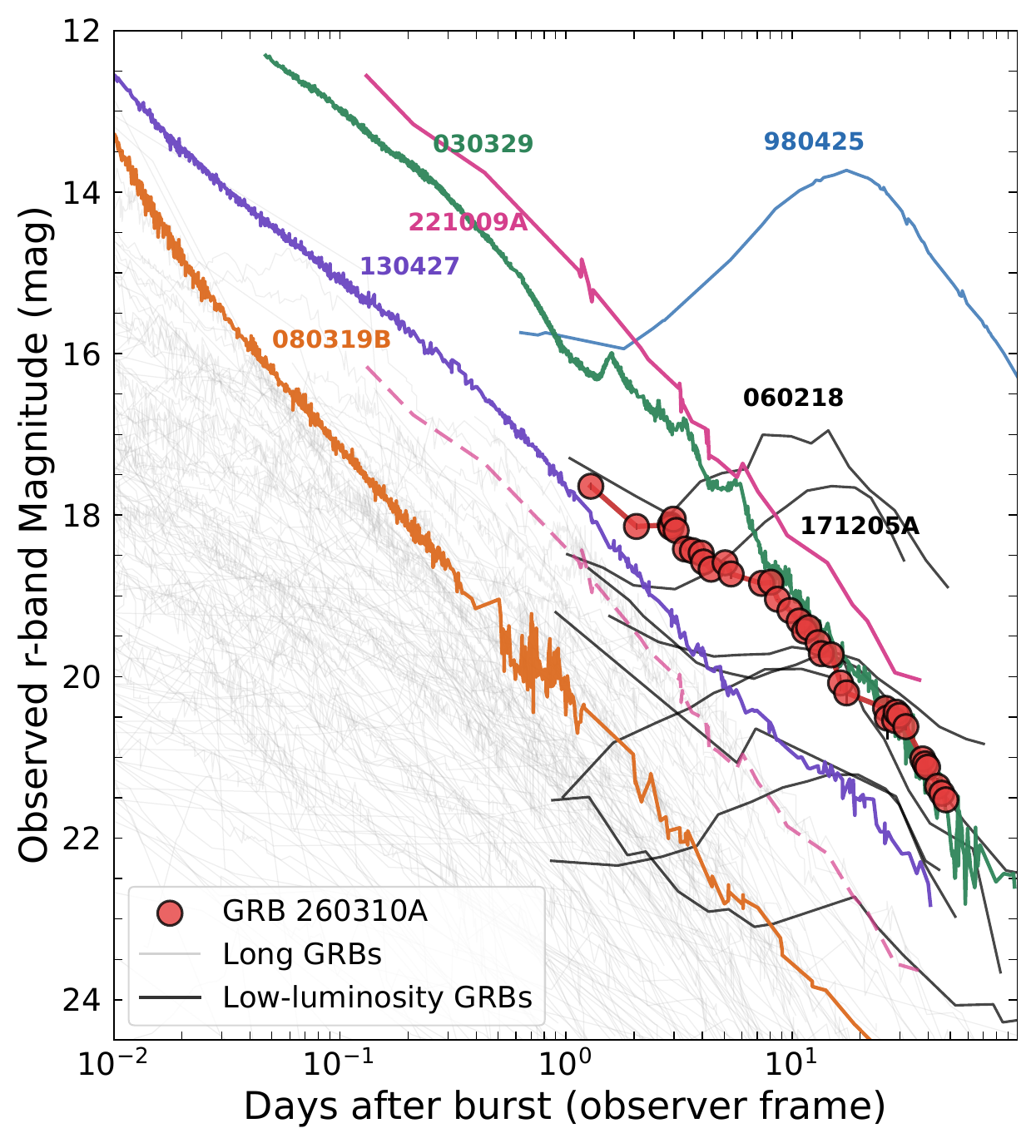}
    \caption{Observed optical lightcurves of long GRBs (gray) and low-luminosity GRBs (black) compared to GRB 260310A (red circles). A few specific GRBs are highlighted for a more specific comparison. GRB 221009A (magenta) is shown both observed (dashed) and dust corrected (solid). Data has been compiled from \citet[][]{Galama1998,McKenzie1999,Sollerman2000, Sollerman2002,Pian2006,Sollerman2006,Kann2006,Kann2010,Kann2011,Clocchiatti2011,Nicuesa2012,Perley2014,DElia2018,OConnor2023}.}
    \label{fig:Kann}
\end{figure}

\section{Discussion}
\label{sec:discuss}

\subsection{Optical Afterglow Properties}

Due to its nearby distance ($z$\,$=$\,$0.153$), GRB 260310A displays an extremely bright multi-wavelength counterpart. At optical wavelengths, it is among the brightest GRB afterglows ever discovered. To demonstrate this, in Figure \ref{fig:Kann}, we show the $r$-band lightcurve of GRB 260310A versus a sample of 190 long GRB afterglows. The optical light curves shown in Figure \ref{fig:Kann} were largely obtained from a public repository\footnote{\url{https://github.com/steveschulze/kann_optical_afterglows}} of GRB light curves initially compiled by David Alexander Kann \citep[see][]{Kann2006,Kann2010,Kann2011,Nicuesa2012} and supplemented by additional events.

Based on this, we find that GRB 260310A displays the brightest observed optical afterglow in the era of the \textit{Neil Gehrels Swift Observatory} \citep{Gehrels2004} at optical wavelengths \citep[see also][]{Yang2026,Gill2026,Hinds2026}. This is largely due to the intervening Galactic dust impacting GRB 221009A \citep{OConnor2023,Laskar2023,Levan2023-221009A}, as its dust corrected brightness exceeds GRB 260310A. Additionally, we have excluded GRBs 980425, 060218, and 171205A, see Figure \ref{fig:Kann}. While these bursts clearly exceed its brightness at optical wavelengths, their optical lightcurves are generally attributed to a combination of shock cooling emission at early times and supernova emission at late times \citep[e.g.,][]{Campana060218,Pian2006,Soderberg2006grb060218,Cano2011,Izzo2019}. They simply exceed the observed brightness of GRB 260310A due to their significantly closer distances. Comparing GRB 260310A to energetic GRBs at a similar redshift ($z$\,$\approx$\,$0.15$), such as GRB 221009A at $z$\,$=$\,$0.151$ \citep{deUgartePostigo2022z,Malesani2025} and GRB 030329 at $z$\,$=$\,$0.1685$ \citep{Hjorth2003,Stanek2003}, we find that it closely matches the observed brightness of GRB 030329, while lying below the dust-corrected lightcurve of GRB 221009A. 

However, even correcting for the redshift, we find that GRB 260310A is among the most luminous GRB afterglows, and comparable to GRB 130427A \citep[$z$\,$=$\,$0.34$;][]{Xu2013,Perley2014,Maselli13,Vestrand14} and GRB 030329 at late times ($\gtrsim$\,$10$ d). At this phase the supernova contribution begins to emerge (Figure \ref{fig:AGSNfit}) and takes over as the primary contributor to the lightcurve. 



\begin{figure}
    \centering
    \includegraphics[width=\linewidth]{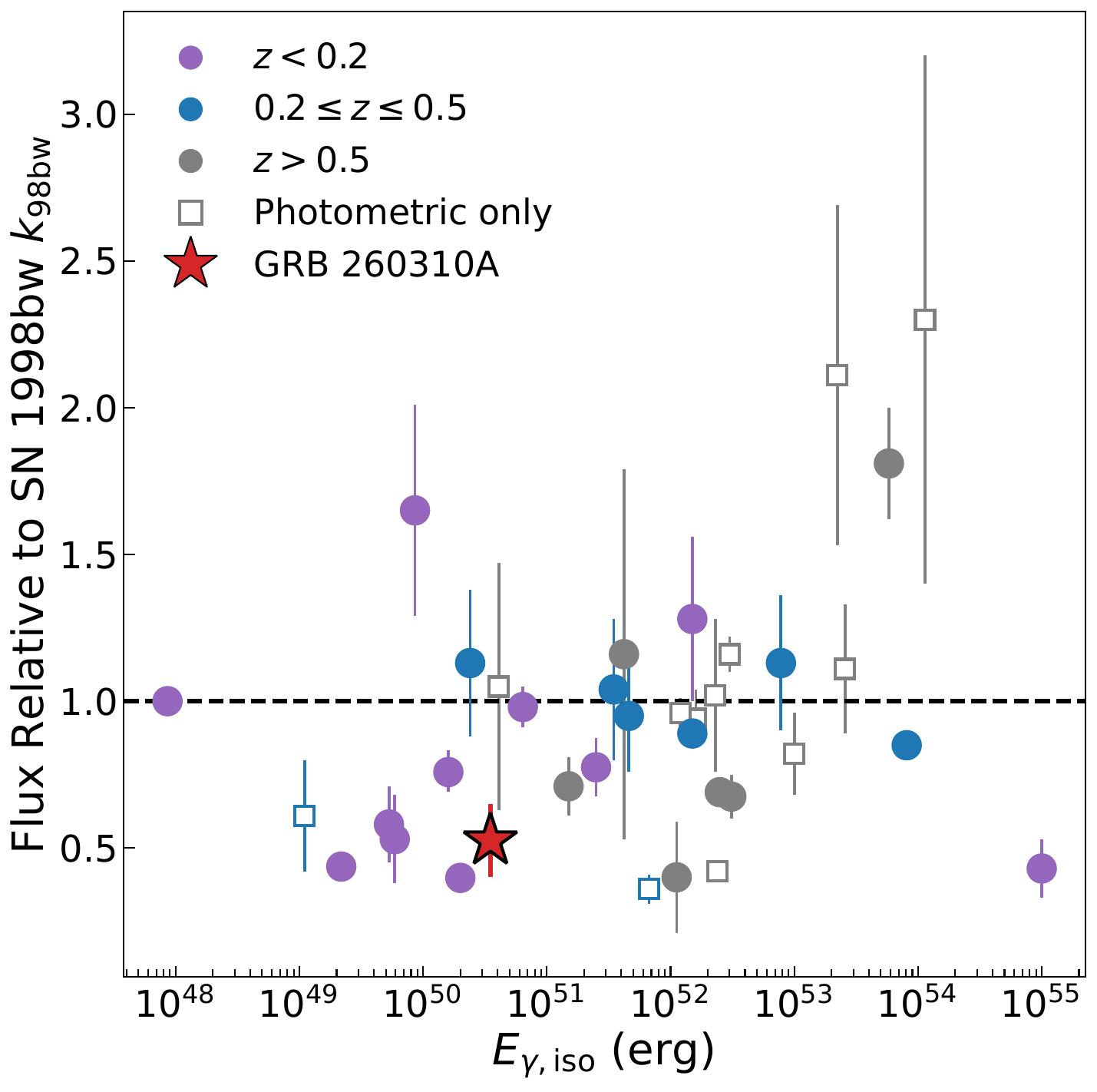}
    \caption{The flux-stretching factor $k_\textrm{98bw}$ relative to SN 1998bw in rest-frame $V$-band of a sample of GRB-SNe is shown versus the isotropic-equivalent gamma-ray energy $E_{\gamma,\textrm{iso}}$ release of the GRB prompt emission. The data was compiled from \citet{Hjorth2012sn,Hjorth2013,Greiner2015,Cano2017,Cano2016jca,Klose2019,Izzo2019,Hu2021,Srinivasaragavan2023,Hu2021,Rossi2026sn} and references therein. This figure is reproduced from \citet{Hjorth2013} and \citet{Srinivasaragavan2023}.
    }
    \label{fig:k98bw-Eiso}
\end{figure}

\subsection{Comparison to Other GRB-SN}



Here we compare the properties of GRB 260310A/SN 2026fgk to other Type Ic-BL GRB-SN identified in the literature. We compiled and cross-checked the nearby GRB-SN sample by combining multiple GRB-SN catalogs \citep{Hjorth2012sn,Cano2017,GRBSNtool} and adding more recent secure additions to the literature \citep[e.g.,][]{Klose2019}. In total, there are 73 GRB-SN that have been identified in the literature \citep[see, e.g.,][and references therein]{GRBSNtool}, of which 38 have spectroscopic classifications of varying signal-to-noise depending largely on the burst's redshift and afterglow brightness. Within $z$\,$<$\,$0.2$, there are 11 previous GRB SN, making GRB 260310A the 12th case of a low-z GRB-SN with spectroscopic confirmation. These events are GRB 980425/SN 1998bw \citep{Galama1998,Stathakis2000,Patat2001}, GRB 030329/SN 2003dh \citep{Hjorth2003,Stanek2003,Matheson2003}, GRB 031203/SN 2003lw \citep{Malesani2004}, GRB 060218/SN 2006aj \citep{Campana060218,Pian2006,Soderberg2006grb060218,Sollerman2006,Ferrero2006,Mazzali2006}, GRB 100316D/SN 2010bh \citep{Chornock2010,Starling2011,Cano2011,Fan2011,Bufano2012,Olivares2012}, GRB 130702A/SN 2013dx \citep{DElia2015,Toy2016,Volnova2017,Mazzali2021}, GRB 161219B/SN 2016jca \citep{Cano2016jca,Ashall2019}, GRB 171205A/SN 2017iuk \citep{DElia2018,Wang2018sn,Izzo2019}, GRB 180728A/SN 2018fip \citep{Rossi2026sn}, GRB 190829A/SN 2019oyw \citep{Hu2021,Bhirombhakdi2024}, and GRB 221009A/SN 2022xiw \citep{Srinivasaragavan2023,Fulton2023, Shrestha2023,Blanchard2024}. Since the launch of the Einstein Probe \citep[EP;][]{Yuan2025}, there have been an additional 4 Type Ic-BL supernova associated with high-energy (X-ray) detections at $z$\,$<$\,$0.2$: EP250108a/SN 2025kg \citep{Srinivasaragavan2025EP0108a,Eyles-Ferris2025EP,Rastinejad2025EP,Li2025}, EP250304A/ SN 2025fhm \citep{EP250304a-SN-GCN,Cotter2026}, EP250827b/SN 2025wkm \citep{Srinivasaragavan2026}, and, occurring after the discovery of GRB 260310A, EP260321a/SN 2026gzf \citep[][]{Corcoran2026gcn,OConnor2026,Yuan2026,Rastinejad2026,MartinCarrillo2026,Chen2026}. 

In Figure \ref{fig:k98bw-Eiso}, we compare the flux-stretching factor $k_\textrm{98bw}$ of these GRB-SN relative to the canonical SN 1998bw \citep{Galama1998} versus the isotropic-equivalent gamma-ray energy release of the GRB's prompt emission. While it is well known that there is no significant correlation between the supernova brightness and the GRB prompt emission \citep[e.g.,][]{Hjorth2013,Schulze2014,Srinivasaragavan2023}, there is an obvious selection effect of identifying the faintest explosions, of both types, at low redshifts ($z$\,$<$\,$0.2$). We find that GRB 260310A matches well with the distribution of other GRB-SN \citep[see also][]{Gill2026,Hinds2026}.

Additionally, our inferred supernova ejecta mass of $4$\,$-$\,$6$\,$M_\odot$ with nickel mass $\sim$\,$0.4$\,$-$\,$0.5$\,$M_\odot$ and total kinetic energy $E_\textrm{k}$\,$=$\,$(3-8)\times10^{51}$ erg align well with other GRB-SN. For example, the analysis of \citet{Cano2017} presents ``average'' GRB-SN properties of $E_\textrm{k}$\,$=$\,$(2.5\pm1.8)\times10^{52}$ erg, $M_\textrm{ej}$\,$=$\,$6\pm4$\,$M_\odot$, and $M_\textrm{Ni}$\,$=$\,$0.4\pm0.2$\,$M_\odot$. This is consistent with the agreement between the SN 1998bw flux-stretching inferred for SN 2026fgk as compared to other GRB-SN (see Figure \ref{fig:k98bw-Eiso}).

\begin{figure}
    \centering
    \includegraphics[width=\linewidth]{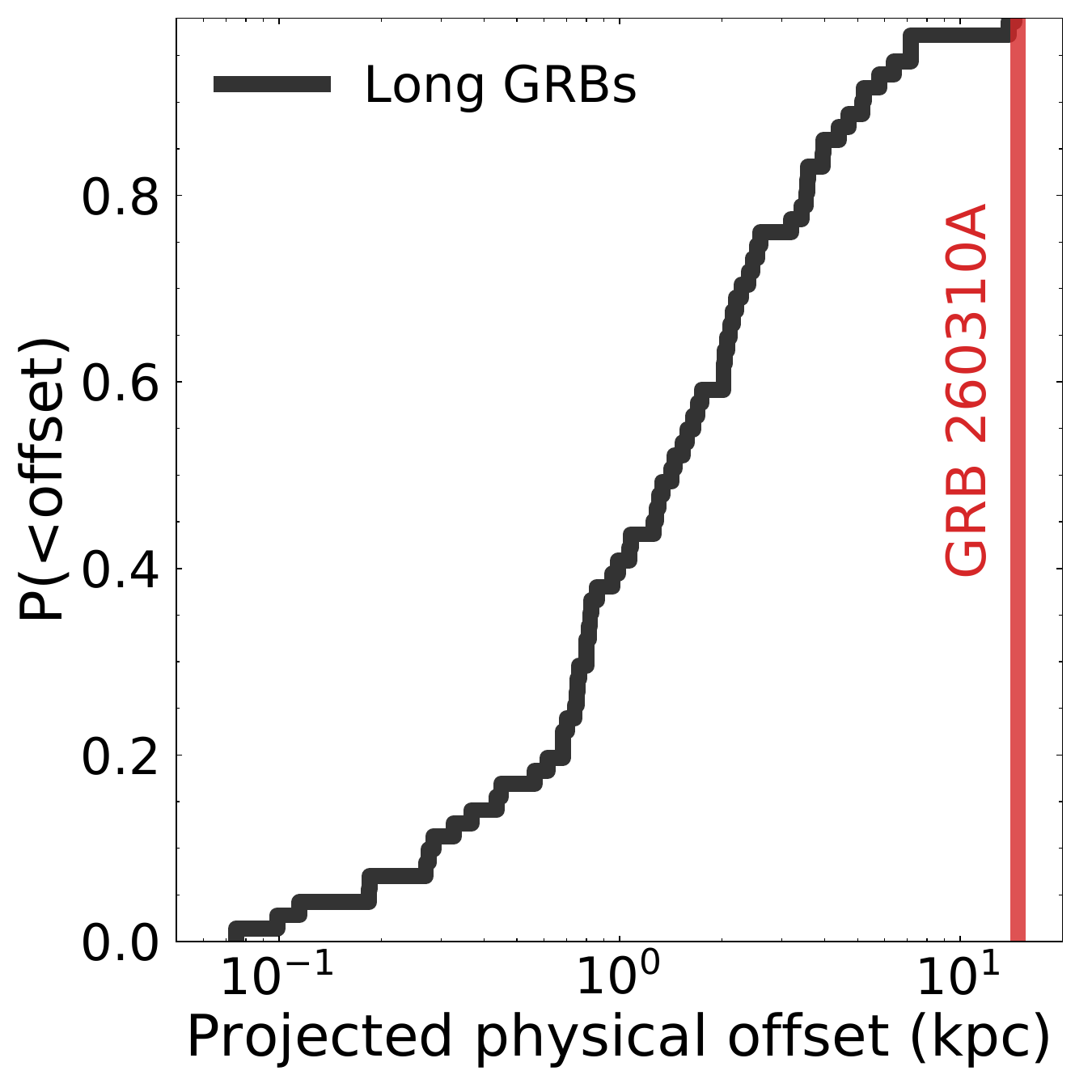}
    \caption{Cumulative distribution of projected physical offsets for long GRBs from their host galaxies \citep{Blanchard2016,Lyman2017}. The location of GRB 260310A is shown as a red vertical line.
    } 
    \label{fig:offset}
\end{figure}

\begin{figure}
    \centering
    \includegraphics[width=\columnwidth]{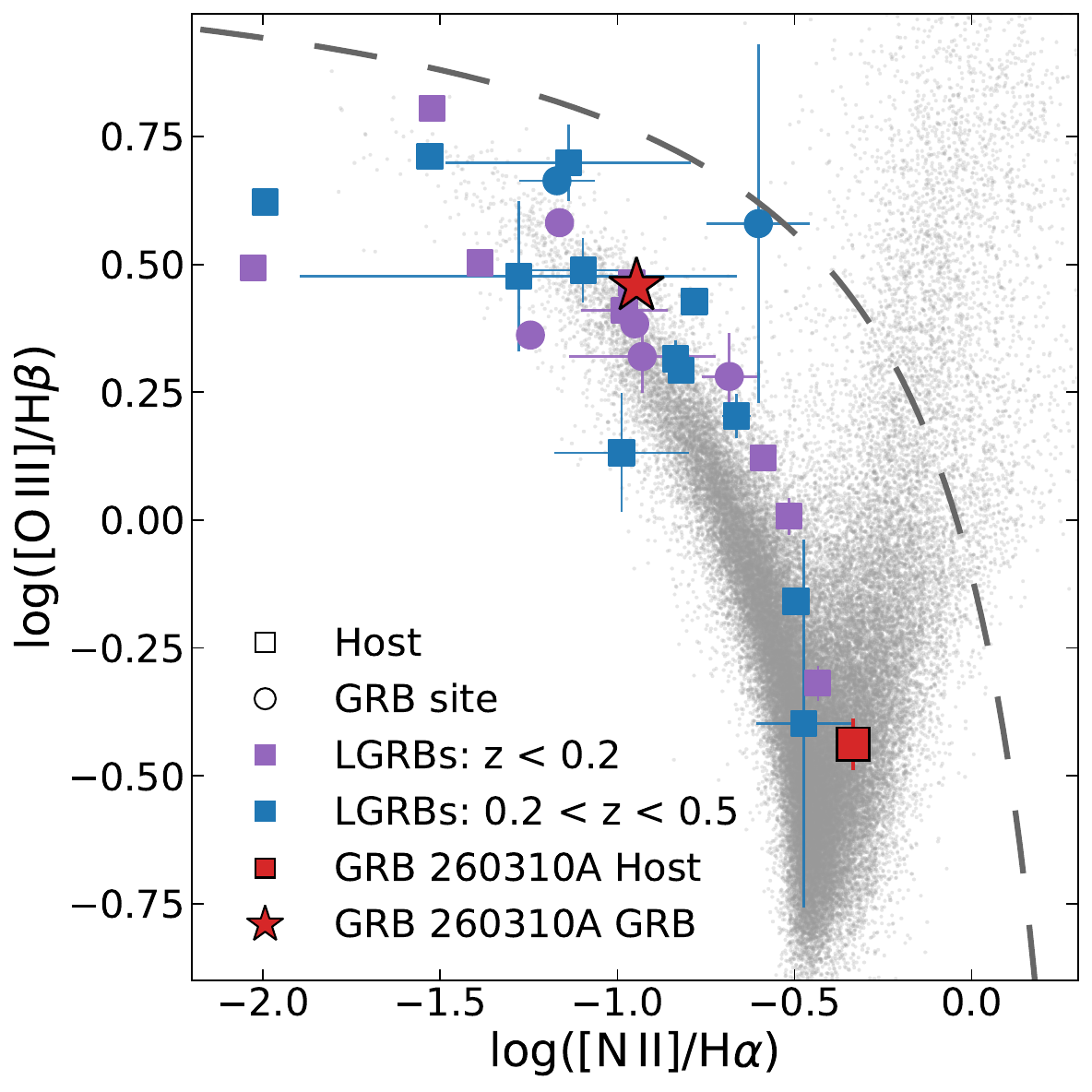}
    \caption{BPT diagram for long GRB host galaxies (squares) and explosion sites (circles) compared to a background population of SDSS galaxies (small gray circles; \citealt{SDSS,SDSS-DR8,2019MNRAS.485.1085S}). Both the explosion site (red star) and integrated host galaxy (red square) of GRB 260310A are shown. The long GRB data is compiled from \citet{Christensen2008,Han2010,Kelly2013,Schulze2014,Stanway2015,Kruhler2015,Izzo2017,Cano2016jca,Heintz2018,deUgartePostigo2018,Melandri2019,deUgartePostigo2020,Thone2024}.
 }
    \label{fig:bpt}
\end{figure}

\subsection{Local Environment and Galactocentric Offset}







Long GRBs are understood to arise from the deaths of massive stars \citep{Woosley1993,Woosley2006} and are therefore expected to trace recent star formation. Consistent with that picture, population studies have shown that long GRBs are typically concentrated on the brightest ultraviolet light of their hosts at small galactocentric offsets \citep{Bloom2002,Fruchter2006,Blanchard2016,Lyman2017}. GRB 260310A is unusual in this context because its projected separation from its host galaxy is large, $5.4\pm0.2\arcsec$, corresponding to a projected physical offset of $14.8\pm0.6$ kpc at $z$\,$=$\,$0.1530$ (Figure \ref{fig:GemFC}). This is $10\times$ larger than the median value for long GRBs \citep{Blanchard2016}, see Figure \ref{fig:offset}). However, the host normalized offset is only $R/R_\textrm{e}$\,$=$\,$2.3\pm0.1$, where $R_\textrm{e}$\,$=$\,$2.4\arcsec$ is the half-light radius of the host galaxy. Moreover, the Gemini spectra demonstrate that the burst is not dynamically detached from the host (Figure \ref{fig:Gem2D}). Instead, the redshift difference between the GRB site ($z$\,$=$\,$0.153$) and the host ($z$\,$=$\,$0.1525$) is only $\sim 100$ km s$^{-1}$, nebular emission forms a clear bridge between the galaxy and the burst position, and H$\alpha$ emission extends to comparable distances on the opposite side of the galaxy (Figure \ref{fig:Gem2D}). Taken together, these observations indicate that GRB 260310A occurred in an outlying H \textsc{ii} region embedded within an extended, low surface brightness star forming disk, rather than in an isolated satellite or halo environment. In that sense, the event appears to belong to the same broader class of long GRBs with large galactocentric offsets ($>$\,$5$ kpc) but still clearly locally star forming environments identified in other low redshift systems \citep{Christensen2008,Levesque2011,Schulze2014,Izzo2017,Melandri2019,Thone2024}.

The GRB site itself remains actively star forming, with ${\rm SFR}_{\rm H\alpha}$\,$\approx$\,$1\,M_\odot\,{\rm yr}^{-1}$, and is clearly less metal enriched than the host galaxy ($12+\log({\rm O/H})\approx8.6$), with a sub-solar metallicity $12+\log({\rm O/H})\approx8.3$ (see \S \ref{sec:host}). This is consistent with the expectations for a collapsar progenitor and with observations of other GRB environments \citep[e.g.,][]{Levesque2010-1,Levesque2010-2,Kruhler2015}. Thus, while the burst occurred at large projected radius, it has a locally favorable environment for long GRB production, namely a star forming region with lower metallicity than the galaxy integrated value (Figure \ref{fig:bpt}). That result is fully consistent with the resolved view of nearby long GRB hosts, in which the immediate burst environment is often less enriched than the host as a whole even when the galaxy integrated spectrum appears comparatively evolved \citep{Levesque2011,Thone2014,Kruhler2015,Kruhler2017,Izzo2017,Cano2016jca,Tanga2018,Michalowski2018,Melandri2019,Thone2021,Thone2024}. GRB 260310A therefore illustrates an important point about the interpretation of GRB host environments. Large projected offsets do not necessarily imply that the progenitor formed in a remote or quiescent region of the host galaxy. In this case, our Gemini spectroscopy definitively shows that the burst site is part of a continuous star forming structure connected to the main body of the galaxy, and that the metallicity declines toward that outer region approaching the GRB explosion site (Figure \ref{fig:Gem2D}).

\section{Conclusions}
\label{sec:conclusions}

We present the results of our comprehensive optical and near-infrared observing campaign of GRB 260310A utilizing FTW, Seimei, DESI, HET, and Gemini. At the nearby distance of $z$\,$=$\,$0.153$ (756 Mpc), GRB 260310A provided a critical opportunity to search for and identify a supernova component. Our investigation yielded the spectroscopic classification of the associated SN 2026fgk, confirming GRB 260310A's relation to other long GRBs with collapsar progenitors. SN 2026fgk is only the 12th GRB-SN uncovered within 1 Gpc ($z$\,$<$\,$0.2$) since 1998. It is the 15th Ic-BL supernova within 1 Gpc associated with a prompt high-energy transient when including recent EP fast X-ray transients and adds to the growing sample of spectroscopically classified GRB-SN. 

We find that the optical afterglow of GRB 260310A is among the brightest ever observed, matching GRB 030329 at late-times and exceeding the observed optical lightcurve of GRB 221009A. The brightness of the afterglow, combined with the rather smooth onset of supernova features, masked the supernova discovery and it was not robustly uncovered spectrally until $\sim17$ d post-trigger. Only a handful of $z$\,$<$\,$0.2$ GRB-SN had supernova spectral features identified later than this, e.g., GRB 031203/SN 2003lw \citep{Malesani2004} and GRB 221009A/SN 2022xiw \citep{Blanchard2024}. 

Additionally, while the $\sim$15 kpc offset of GRB 260310A from its putative host galaxy is among the largest ever observed for a long GRB, we find that this is simply due to an offset H \textsc{ii} region. This H \textsc{ii} region is linked to the host galaxy by a bridge of star formation as observed along the Gemini GMOS slit. The offset is in the direction of the extension of the host galaxy disk, and we find that it is simply be due to an extended low surface brightness disk. The GRB explosion site has a very low metallicity compared to the host galaxy, as expected for a collapsar progenitor and consistent with the explosion sites of other GRB-SN.

GRB 260310A continues the line of low redshift ($z$\,$<$\,$0.2$) GRBs with outlier properties in both prompt emission and galactocentric offset, and demonstrates the diversity in GRB explosion properties. Due to its nearby distance, and circumpolar sky position, additional observations of both the GRB-SN and afterglow to significantly later times are both possible and warranted to improve constraints on the physical parameters of the explosion. We strongly encourage additional follow-up at all wavelengths.

\begin{acknowledgments}

The authors thank the referee for providing useful comments and suggestions.

BO thanks the Gemini Director Time Review Board (GDTRB), including  Aleksandar Cikota, Hwihyun Kim, Bryan Miller, Siyi Xu, and Elena Sabbi, for approving the DDT request. BO is grateful for the excellent support of Atsuko Nitta, Teo Mocnik, Jen Andrews, Sunny Stewart, Denise Hung, 
Kristin Chiboucas, Hyewon Suh and the entire Gemini-North staff in rapidly scheduling and obtaining these observations. 

BO is supported by the McWilliams Postdoctoral Fellowship in the McWilliams Center for Cosmology and Astrophysics at Carnegie Mellon University. 

This paper contains data obtained at the Wendelstein Observatory of the Ludwig-Maximilians University Munich. Funded by the Deutsche Forschungsgemeinschaft (DFG, German Research Foundation) under Germany's Excellence Strategy – EXC-2094/2 – 390783311.
The authors acknowledge the IT Support Team of the Faculty of Physics at LMU for maintaining and supporting the HPC, storage, and computational infrastructure used in this work.

This material is based upon work supported by the U.S. Department of Energy (DOE), Office of Science, Office of High-Energy Physics, under Contract No. DE–AC02–05CH11231, and by the National Energy Research Scientific Computing Center, a DOE Office of Science User Facility under the same contract. Additional support for DESI was provided by the U.S. National Science Foundation (NSF), Division of Astronomical Sciences under Contract No. AST-0950945 to the NSF’s National Optical-Infrared Astronomy Research Laboratory; the Science and Technology Facilities Council of the United Kingdom; the Gordon and Betty Moore Foundation; the Heising-Simons Foundation; the French Alternative Energies and Atomic Energy Commission (CEA); the National Council of Humanities, Science and Technology of Mexico (CONAHCYT); the Ministry of Science, Innovation and Universities of Spain (MICIU/AEI/10.13039/501100011033), and by the DESI Member Institutions: \url{https://www.desi.lbl.gov/collaborating-institutions}. The authors are honored to be permitted to conduct scientific research on I'oligam Du'ag (Kitt Peak), a mountain with particular significance to the Tohono O’odham Nation. Any opinions, findings, and conclusions or recommendations expressed in this material are those of the author(s) and do not necessarily reflect the views of the U. S. National Science Foundation, the U. S. Department of Energy, or any of the listed funding agencies.

Based on observations obtained at the international Gemini Observatory, a program of NSF's OIR Lab, which is managed by the Association of Universities for Research in Astronomy (AURA) under a cooperative agreement with the National Science Foundation on behalf of the Gemini Observatory partnership: the National Science Foundation (United States), National Research Council (Canada), Agencia Nacional de Investigaci\'{o}n y Desarrollo (Chile), Ministerio de Ciencia, Tecnolog\'{i}a e Innovaci\'{o}n (Argentina), Minist\'{e}rio da Ci\^{e}ncia, Tecnologia, Inova\c{c}\~{o}es e Comunica\c{c}\~{o}es (Brazil), and Korea Astronomy and Space Science Institute (Republic of Korea). The data were acquired through the Gemini Observatory Archive at NSF NOIRLab and processed using DRAGONS (Data Reduction for Astronomy from Gemini Observatory North and South). The authors wish to recognize and acknowledge the very significant cultural role and reverence that the summit of Maunakea has always had within the indigenous Hawaiian community. 

This paper contains data from observations obtained with the Hobby-Eberly Telescope (HET), which is a joint project of the University of Texas at Austin, the Pennsylvania State University, Ludwig-Maximillians-Universität München, and Georg-August Universität Göttingen. The HET is named in honor of its principal benefactors, William P. Hobby and Robert E. Eberly. We acknowledge the Texas Advanced Computing Center (TACC) at The University of Texas at Austin for providing high-performance computing, visualization, and storage resources that have contributed to the results reported within this paper. The Low Resolution Spectrograph 2 (LRS2) was developed and funded by the University of Texas at Austin, McDonald Observatory, Department of Astronomy, and Pennsylvania State University. We thank the Leibniz-Institut für Astrophysik Potsdam (AIP) and the Institut für Astrophysik Göttingen (IAG) for their contributions to the construction of the integral field units.

The authors thank the TriCCS developer team (which has been supported by the JSPS KAKENHI grant Nos. JP18H05223, JP20H00174, and JP20H04736, and by NAOJ Joint Development Research).

This research has made use of the GRBSN webtool \citep{GRBSNtool}, available at https://grbsn.watchertelescope.ie.


\end{acknowledgments}





%
\facilities{WO:2m, Seimei, Gemini-North, Mayall, HET
}

\software{\texttt{Astropy} \citep{2018AJ....156..123A,2022ApJ...935..167A}, \texttt{SFFT} \citep{hu_image_2022}, \texttt{AstrOmatic} \citep{Bertin1996,Bertin2006,Bertin2010,2002ASPC..281..228B}, \texttt{DRAGONS} \citep{Labrie2019,Labrie2023}, \texttt{pPXF} \citep{ppxf,ppxf2,ppxf3}, \texttt{redback} \citep{redback}, \texttt{Bilby} \citep{bilby}, \texttt{dynesty} \citep{dynesty}, \texttt{sncosmo} \citep{sncosmo}
}


\appendix

\section{Log of Observations}

In Tables \ref{tab:grb260310a_photometry} and \ref{tab:grb260310a_spectra}, we present a log of all photometric and spectroscopic observations obtained and analyzed in this work.

\begin{longtable}{llllll}
\caption{Log of photometry of GRB 260310A from FTW and Seimei. Times $\Delta T$ are given relative to the GRB trigger. The photometry has been corrected for Galactic reddening $E(B-V)=0.02$ mag \citep{Schlafly2011}.}\label{tab:grb260310a_photometry}\\
\hline
\textbf{Start Time (UT)} & \textbf{$\Delta T$ (d)} & \textbf{Exposure (s)} & \textbf{Telescope} & \textbf{Filter} & \textbf{AB magnitude} \\
\hline
\endfirsthead

\hline
\textbf{Start Time (UT)} & \textbf{$\Delta T$ (d)} & \textbf{Exposure (s)} & \textbf{Telescope} & \textbf{Filter} & \textbf{AB magnitude} \\
\hline
\endhead

\hline
\endfoot

\hline
\endlastfoot
	2026-03-13 19:37:11 & 3.61 & 2160 & FTW & $g$ & $18.66 \pm 0.01$ \\
	2026-03-13 19:37:11 & 3.61 & 2160 & FTW & $i$ & $18.16 \pm 0.01$ \\
	2026-03-13 19:37:28 & 3.61 & 2037 & FTW & $J$ & $17.50 \pm 0.01$ \\
	2026-03-13 19:48:45 & 3.62 & 2400 & FTW & $r$ & $18.39 \pm 0.01$ \\
	2026-03-13 19:48:45 & 3.62 & 1950 & FTW & $z$ & $17.94 \pm 0.02$ \\
	2026-03-13 19:49:01 & 3.62 & 1432 & FTW & $K_s$ & $17.07 \pm 0.01$ \\
	2026-03-14 14:05:06 & 4.38 & 210 & Seimei & $r$ & $18.62 \pm 0.04$ \\
	2026-03-14 14:05:06 & 4.38 & 210 & Seimei & $g$ & $18.87 \pm 0.03$ \\
	2026-03-14 14:05:06 & 4.38 & 90 & Seimei & $i$ & $18.35 \pm 0.04$ \\
	2026-03-14 14:07:44 & 4.38 & 120 & Seimei & $z$ & $18.11 \pm 0.04$ \\
	2026-03-15 13:50:50 & 5.37 & 180 & Seimei & $r$ & $18.68 \pm 0.06$ \\
	2026-03-15 13:50:50 & 5.37 & 180 & Seimei & $g$ & $18.86 \pm 0.06$ \\
	2026-03-15 13:50:50 & 5.37 & 90 & Seimei & $i$ & $18.44 \pm 0.05$ \\
	2026-03-15 13:53:25 & 5.37 & 90 & Seimei & $z$ & $18.18 \pm 0.05$ \\
	2026-03-17 12:44:25 & 7.32 & 390 & Seimei & $r$ & $18.80 \pm 0.02$ \\
	2026-03-17 12:44:26 & 7.32 & 300 & Seimei & $z$ & $18.45 \pm 0.02$ \\
	2026-03-17 12:44:26 & 7.32 & 390 & Seimei & $g$ & $19.10 \pm 0.02$ \\
	2026-03-17 12:50:20 & 7.33 & 90 & Seimei & $i$ & $18.63 \pm 0.03$ \\
	2026-03-18 19:20:28 & 8.60 & 1080 & FTW & $g$ & $19.32 \pm 0.01$ \\
	2026-03-18 19:20:28 & 8.60 & 1080 & FTW & $i$ & $18.83 \pm 0.01$ \\
	2026-03-18 19:20:43 & 8.60 & 1018 & FTW & $J$ & $18.21 \pm 0.02$ \\
	2026-03-18 19:33:26 & 8.61 & 1200 & FTW & $r$ & $19.00 \pm 0.01$ \\
	2026-03-18 19:33:26 & 8.61 & 1200 & FTW & $z$ & $18.68 \pm 0.02$ \\
	2026-03-18 19:33:41 & 8.61 & 849 & FTW & $K_s$ & $17.77 \pm 0.02$ \\
	2026-03-19 23:22:38 & 9.77 & 540 & FTW & $g$ & $19.46 \pm 0.01$ \\
	2026-03-19 23:22:38 & 9.77 & 540 & FTW & $i$ & $18.98 \pm 0.01$ \\
	2026-03-19 23:22:54 & 9.77 & 509 & FTW & $J$ & $18.38 \pm 0.02$ \\
	2026-03-19 23:35:29 & 9.78 & 600 & FTW & $r$ & $19.14 \pm 0.01$ \\
	2026-03-19 23:35:29 & 9.78 & 600 & FTW & $z$ & $18.88 \pm 0.02$ \\
	2026-03-19 23:35:45 & 9.78 & 424 & FTW & $K_s$ & $17.88 \pm 0.03$ \\
	2026-03-20 22:36:22 & 10.74 & 540 & FTW & $g$ & $19.58 \pm 0.01$ \\
	2026-03-20 22:36:22 & 10.74 & 540 & FTW & $i$ & $19.07 \pm 0.01$ \\
	2026-03-20 22:36:39 & 10.74 & 509 & FTW & $J$ & $18.55 \pm 0.02$ \\
	2026-03-20 22:47:55 & 10.74 & 600 & FTW & $r$ & $19.27 \pm 0.04$ \\
	2026-03-20 22:47:55 & 10.74 & 600 & FTW & $z$ & $18.94 \pm 0.07$ \\
	2026-03-20 22:48:12 & 10.74 & 265 & FTW & $K_s$ & $18.14 \pm 0.12$ \\
	2026-03-21 13:01:08 & 11.34 & 240 & Seimei & $r$ & $19.39 \pm 0.06$ \\
	2026-03-21 13:01:09 & 11.34 & 240 & Seimei & $g$ & $19.68 \pm 0.05$ \\
	2026-03-21 13:01:08 & 11.34 & 150 & Seimei & $i$ & $19.06 \pm 0.05$ \\
	2026-03-21 13:04:32 & 11.34 & 90 & Seimei & $z$ & $19.07 \pm 0.08$ \\
	2026-03-22 00:30:25 & 11.81 & 540 & FTW & $g$ & $19.67 \pm 0.01$ \\
	2026-03-22 00:30:25 & 11.81 & 540 & FTW & $i$ & $19.21 \pm 0.02$ \\
	2026-03-22 00:30:42 & 11.81 & 509 & FTW & $J$ & $18.64 \pm 0.03$ \\
	2026-03-22 00:41:59 & 11.82 & 600 & FTW & $r$ & $19.35 \pm 0.01$ \\
	2026-03-22 00:41:59 & 11.82 & 600 & FTW & $z$ & $19.16 \pm 0.03$ \\
	2026-03-22 00:42:16 & 11.82 & 424 & FTW & $K_s$ & $18.28 \pm 0.04$ \\
	2026-03-23 15:59:01 & 13.46 & 300 & Seimei & $g$ & $19.96 \pm 0.06$ \\
	2026-03-23 15:59:01 & 13.46 & 300 & Seimei & $r$ & $19.67 \pm 0.06$ \\
	2026-03-23 15:59:02 & 13.46 & 150 & Seimei & $i$ & $19.53 \pm 0.07$ \\
	2026-03-23 16:02:23 & 13.46 & 150 & Seimei & $z$ & $19.41 \pm 0.09$ \\
	2026-03-24 22:17:30 & 14.72 & 1018 & FTW & $J$ & $19.14 \pm 0.04$ \\
	2026-03-24 22:31:10 & 14.73 & 637 & FTW & $K_s$ & $18.98 \pm 0.10$ \\
	2026-03-25 02:03:56 & 14.88 & 540 & FTW & $g$ & $20.16 \pm 0.05$ \\
	2026-03-25 02:03:56 & 14.88 & 540 & FTW & $i$ & $19.60 \pm 0.04$ \\
	2026-03-25 02:16:46 & 14.89 & 600 & FTW & $r$ & $19.68 \pm 0.03$ \\
	2026-03-25 02:16:46 & 14.89 & 600 & FTW & $z$ & $19.55 \pm 0.08$ \\
	2026-03-26 13:51:30 & 16.37 & 900 & Seimei & $r$ & $20.03 \pm 0.10$ \\
	2026-03-26 13:51:30 & 16.37 & 450 & Seimei & $z$ & $19.68 \pm 0.13$ \\
	2026-03-26 13:51:30 & 16.37 & 900 & Seimei & $g$ & $20.45 \pm 0.10$ \\
	2026-03-26 13:54:53 & 16.37 & 450 & Seimei & $i$ & $19.56 \pm 0.08$ \\
	2026-03-27 14:49:36 & 17.41 & 600 & Seimei & $r$ & $20.16 \pm 0.14$ \\
	2026-03-27 14:49:36 & 17.41 & 600 & Seimei & $g$ & $20.57 \pm 0.18$ \\
	2026-03-28 03:57:52 & 17.96 & 339 & FTW & $J$ & $19.39 \pm 0.07$ \\
	2026-03-28 04:09:20 & 17.97 & 424 & FTW & $K_s$ & $19.31 \pm 0.11$ \\
	2026-04-02 13:47:42 & 23.37 & 660 & Seimei & $g$ & $20.75 \pm 0.34$ \\
	2026-04-05 02:22:49 & 25.89 & 1800 & FTW & $i$ & $20.00 \pm 0.03$ \\
	2026-04-05 02:22:49 & 25.89 & 1800 & FTW & $r$ & $20.35 \pm 0.03$ \\
	2026-04-05 02:41:07 & 25.91 & 1800 & FTW & $g$ & $20.80 \pm 0.05$ \\
	2026-04-05 02:41:07 & 25.91 & 1800 & FTW & $z$ & $20.11 \pm 0.06$ \\
	2026-04-05 14:00:50 & 26.38 & 90 & Seimei & $r$ & $20.47 \pm 0.27$ \\
	2026-04-05 14:00:51 & 26.38 & 90 & Seimei & $i$ & $20.10 \pm 0.21$ \\
	2026-04-07 14:58:05 & 28.42 & 240 & Seimei & $i$ & $20.24 \pm 0.12$ \\
	2026-04-07 14:58:05 & 28.42 & 240 & Seimei & $g$ & $20.98 \pm 0.19$ \\
	2026-04-07 14:58:05 & 28.42 & 240 & Seimei & $r$ & $20.50 \pm 0.15$ \\
	2026-04-08 00:39:23 & 28.82 & 2550 & FTW & $i$ & $20.08 \pm 0.02$ \\
	2026-04-08 00:39:23 & 28.82 & 2550 & FTW & $r$ & $20.39 \pm 0.02$ \\
	2026-04-08 00:57:41 & 28.83 & 2700 & FTW & $g$ & $21.00 \pm 0.03$ \\
	2026-04-08 00:57:41 & 28.83 & 2700 & FTW & $z$ & $20.00 \pm 0.04$ \\
	2026-04-08 23:23:35 & 29.77 & 2160 & FTW & $i$ & $20.14 \pm 0.02$ \\
	2026-04-08 23:23:35 & 29.77 & 1800 & FTW & $r$ & $20.44 \pm 0.02$ \\
	2026-04-08 23:41:40 & 29.78 & 1950 & FTW & $g$ & $21.05 \pm 0.03$ \\
	2026-04-08 23:41:40 & 29.78 & 2100 & FTW & $z$ & $20.12 \pm 0.04$ \\
	2026-04-11 00:17:46 & 31.81 & 1080 & FTW & $i$ & $20.25 \pm 0.03$ \\
	2026-04-11 00:39:34 & 31.82 & 720 & FTW & $r$ & $20.57 \pm 0.05$ \\
	2026-04-11 00:39:34 & 31.82 & 2520 & FTW & $z$ & $20.33 \pm 0.06$ \\
	2026-04-11 00:48:48 & 31.83 & 720 & FTW & $g$ & $21.27 \pm 0.06$ \\
	2026-04-11 22:07:43 & 32.72 & 360 & FTW & $i$ & $20.28 \pm 0.07$ \\
	2026-04-16 23:13:52 & 37.76 & 2160 & FTW & $g$ & $21.77 \pm 0.05$ \\
	2026-04-16 23:13:52 & 37.76 & 2160 & FTW & $z$ & $20.59 \pm 0.07$ \\
	2026-04-16 23:57:41 & 37.79 & 1440 & FTW & $i$ & $20.69 \pm 0.04$ \\
	2026-04-16 23:57:41 & 37.79 & 1440 & FTW & $r$ & $20.97 \pm 0.05$ \\
	2026-04-17 22:27:48 & 38.73 & 3600 & FTW & $g$ & $21.79 \pm 0.05$ \\
	2026-04-17 22:27:48 & 38.73 & 3600 & FTW & $z$ & $20.70 \pm 0.06$ \\
	2026-04-17 22:49:30 & 38.74 & 2160 & FTW & $i$ & $20.66 \pm 0.04$ \\
	2026-04-17 22:49:30 & 38.74 & 2160 & FTW & $r$ & $21.03 \pm 0.05$ \\
	2026-04-18 21:31:16 & 39.69 & 3420 & FTW & $g$ & $21.89 \pm 0.07$ \\
	2026-04-18 21:31:16 & 39.69 & 1800 & FTW & $z$ & $20.50 \pm 0.09$ \\
	2026-04-18 21:42:26 & 39.70 & 2160 & FTW & $i$ & $20.83 \pm 0.06$ \\
	2026-04-18 21:42:26 & 39.70 & 1740 & FTW & $r$ & $21.08 \pm 0.08$ \\
	2026-04-23 02:26:02 & 43.90 & 1980 & FTW & $z$ & $20.82 \pm 0.10$ \\
	2026-04-23 02:26:17 & 43.90 & 1984 & FTW & $J$ & $20.26 \pm 0.05$ \\
	2026-04-23 02:51:28 & 43.91 & 720 & FTW & $i$ & $20.98 \pm 0.11$ \\
	2026-04-23 02:51:28 & 43.91 & 720 & FTW & $r$ & $21.31 \pm 0.11$ \\
	2026-04-23 02:51:43 & 43.91 & 637 & FTW & $K_s$ & $19.98 \pm 0.13$ \\
	2026-04-25 01:25:55 & 45.85 & 2160 & FTW & $z$ & $21.00 \pm 0.09$ \\
	2026-04-25 01:26:13 & 45.85 & 1697 & FTW & $J$ & $20.29 \pm 0.15$ \\
	2026-04-25 01:48:10 & 45.87 & 1440 & FTW & $i$ & $21.02 \pm 0.06$ \\
	2026-04-25 01:48:10 & 45.87 & 1440 & FTW & $r$ & $21.40 \pm 0.07$ \\
	2026-04-25 01:48:28 & 45.87 & 902 & FTW & $K_s$ & $19.96 \pm 0.20$ \\
	2026-04-25 02:08:33 & 45.88 & 1080 & FTW & $g$ & $22.13 \pm 0.09$ \\
	2026-04-26 22:22:55 & 47.73 & 3395 & FTW & $J$ & $20.33 \pm 0.03$ \\
	2026-04-26 22:41:41 & 47.74 & 1538 & FTW & $K_s$ & $19.95 \pm 0.11$ \\
	2026-04-26 23:23:11 & 47.77 & 1800 & FTW & $g$ & $22.11 \pm 0.11$ \\
	2026-04-26 23:23:11 & 47.77 & 1800 & FTW & $z$ & $21.07 \pm 0.09$ \\
	2026-04-26 23:52:11 & 47.79 & 1440 & FTW & $i$ & $21.13 \pm 0.07$ \\
	2026-04-26 23:52:11 & 47.79 & 1440 & FTW & $r$ & $21.48 \pm 0.09$ \\
\end{longtable}

\begin{table*}
\centering
\caption{Log of spectroscopic observations of GRB 260310A/SN 2026fgk.}
\label{tab:grb260310a_spectra}
\begin{tabular}{lcllll}
\hline
Start Date (UTC) & $\Delta T$ (days) & Telescope & Instrument & Grating & Exposure (s) \\
\hline
2026-03-14 14:24:00 &  4.39 & Seimei      & KOOLS-IFU        & VPH-Blue           & 2550 \\
2026-03-15 14:09:36 &  5.38 & Seimei      & KOOLS-IFU        & VPH-Red            & 2700 \\

2026-03-16 08:53:08 &  6.16 & Mayall        & DESI             & \nodata           &  1200 \\
2026-03-17 08:07:32 &  7.13 & Mayall        & DESI             & \nodata           &  1200 \\
2026-03-18 06:49:15 &  8.08 & Mayall        & DESI             & \nodata           &  1200 \\
2026-03-19 09:50:00 &  9.20 & HET           & LRS2           & LRS2-R       & 1800 \\
2026-03-22 06:59:22 & 12.08 & Mayall       & DESI             & \nodata           &  1200 \\
2026-03-27 14:56:32 & 17.42 & Gemini-North      & GMOS           & B480      & 2120 \\
2026-04-05 12:40:09 & 26.32 & Gemini-North      & GMOS           & B480      & 4000 \\
2026-04-20 13:09:59 & 41.34 & Gemini-North      & GMOS           & B480      & 5901 \\
2026-04-23 12:02:23 & 44.30 & Gemini-North      & GMOS           & B480      & 6000 \\
2026-05-13 07:43:05 & 64.12 & Gemini-North      & GMOS           & B480      & 4000 \\
\hline
\end{tabular}
\end{table*}

\section{Additional Model Fitting Results}
\label{sec:modelfitappendix}

\begin{figure*}
    \centering
    \includegraphics[width=\linewidth]{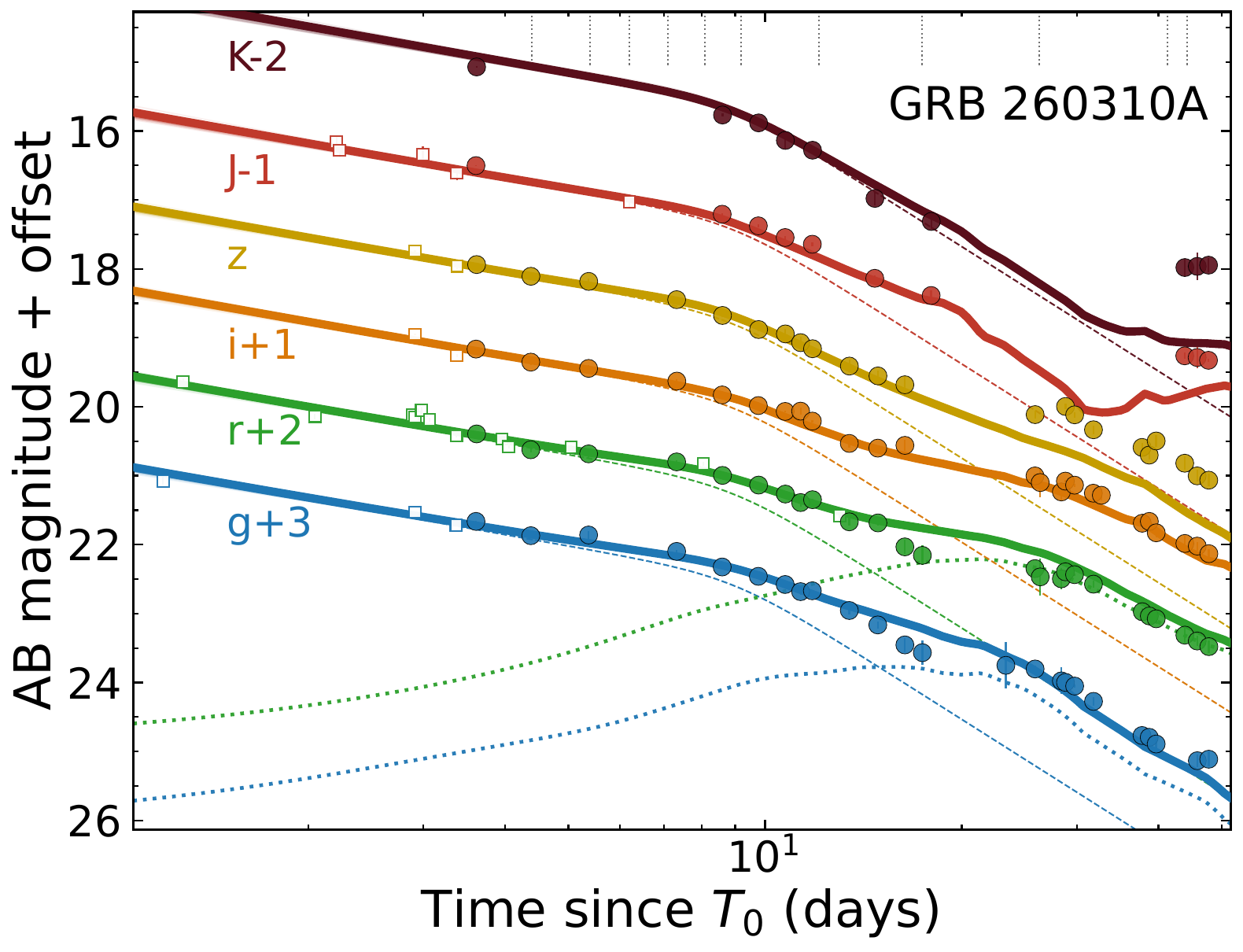}
    \caption{Best-fit model (powerlaw+SN 1998bw) to the multi-wavelength lightcurve of GRB 260310A in the observer-frame. The thick solid lines in each filter represent the total model. The thin dashed lines represent the broken power-law afterglow component and the dotted lines, shown only in the $g$- and $r$-bands to avoid clutter, represent the model lightcurve of SN 1998bw with a flux scaling factor $k_\textrm{98bw}$\,$=$\,$0.50$. Photometry presented in this work is shown as a solid circle and publicly reported photometry compiled from GCN Circulars is shown as empty squares \citep{HindsDisco,GCN1,GCN2,GCN3,GCN4,GCN5,GCN6,GCN7,GCN8,GCN9,GCN10,GCN11,GCN12}.
    }
    \label{fig:AGSNfit-2}
\end{figure*}

In Figure \ref{fig:AGSNfit-2} we show the best-fit model to the multi-wavelength lightcurve for the powerlaw+SN 1998bw model with a SN 1998bw flux scaling factor of $k_\textrm{98bw}$\,$=$\,$0.50$. The use of SN 1998bw as a template is the standard adopted method for such analyses in the literature \citep[e.g.,][]{Bloom1999-980326,Zeh2004}.
We note that the near-infrared ($JK_s$) behavior of the model is uncertain and due to extrapolation to wavelengths beyond the available templates for SN 1998bw \citep[e.g., \texttt{v19-1998bw};][]{Vincenzi2019}. The $JK_s$ model lightcurves of SN 1998bw should therefore be interpreted with caution.

\bibliography{bib}{}
\bibliographystyle{aasjournal}



\end{document}